\documentclass[%
reprint,
superscriptaddress,
 amsmath,amssymb,
 aps,
 pre,
]{revtex4-1}

\usepackage{graphicx}
\graphicspath{{figures/}}
\usepackage{amsmath,amssymb,amsfonts}
\usepackage{subfigure}
\usepackage[maxfloats=100]{morefloats}
\usepackage{dcolumn}
\usepackage{bm}
\usepackage{url}
\usepackage{hyperref}
\newcommand{\R}{\mathbb{R}}
\newcommand{\mywa}{0.15\textwidth}

\newcommand{\mywd}{0.45\textwidth}
\newcommand{\mywe}{0.5\textwidth}

\begin{document}


\title{Extreme learning machine for reduced order modeling of turbulent geophysical flows}

\author{Omer San}
\email{osan@okstate.edu}

\author{Romit Maulik}%
\affiliation{%
 School of Mechanical and Aerospace Engineering, Oklahoma State University, Stillwater, OK 74078, USA\\
}%




\date{\today}

\begin{abstract}
We investigate the application of artificial neural networks to stabilize proper orthogonal decomposition based reduced order models for quasi-stationary geophysical turbulent flows. An extreme learning machine concept is introduced for computing an eddy-viscosity closure dynamically to incorporate the effects of the truncated modes. We consider a four-gyre wind-driven ocean circulation problem as our prototype setting to assess the performance of the proposed data-driven approach. Our framework provides a significant reduction in computational time and effectively retains the dynamics of the full-order model during the forward simulation period beyond the training data set. Furthermore, we show that the method is robust for larger choices of time steps and can be used as an efficient and reliable tool for long time integration of general circulation models.

\end{abstract}

\maketitle


\section{Introduction}
\label{sec:intro}

The spatiotemporal complexity of many applications in the computational sciences leads to very large-scale dynamical systems whose simulations make overwhelming and unmanageable demands on computational resources. Indeed, many problems remain intractable when multiple forward full-order numerical simulations are required. Since the computational cost of these high fidelity simulations is prohibitive, model order reduction approaches, also known as reduced order models (or ROMs), are commonly used to reduce this computational burden in many applications (e.g., see \cite{brunton2015closed} for a review of closed-loop control applications in fluid turbulence, and \cite{daescu2007efficiency,cao2007reduced,daescu2008dual} for a discussion of variational data assimilation applications in weather and climate modeling). A number of recent review articles have addressed the strengths of several modal analysis, reduced basis and model reduction techniques \cite{benner2015survey,rowley2017model,taira2017modal}.
In their survey, dedicated primarily to the reduced order modeling for fluid analysis and control, Rowley and Dawson \cite{rowley2017model} have discussed several techniques including proper orthogonal decomposition (POD), balanced truncation and balanced POD, eigensystem realization algorithms (ERA), dynamic mode decomposition (DMD) and Koopman operator theory with attention devoted to the similarities and analogies between these methods. An excellent overview and introduction to such techniques may also be found in \cite{taira2017modal}.

In this study, we consider the POD framework in combination with the Galerkin projection procedure \cite{holmes1998turbulence}, which is a prominent approach for generating ROMs for nonlinear systems \cite{ito1998reduced,iollo2000stability,noack2003hierarchy}.
The POD procedure identifies the most energetic modes (usually from high-fidelity experimental or numerical data), which are expected to contain the dominant statistical characteristics of these systems. It is therefore possible to provide accurate approximations to the high-fidelity data with a few POD modes in which fine scale details are embedded. The resulting dynamical systems are low dimensional (due to truncation) but dense and provide robust surrogate models for forward simulations. It has been widely used in various disciplines under a variety of different names (e.g., see \cite{narasimha2011kosambi} for an excellent historical discussion).

Although the standard Galerkin projection provides a standardized way to build ROMs, its applicability to complex systems is limited primarily due to modeling errors associated with the truncation of POD modes. The limitation is more prominent in turbulent flow systems where an intense scale separation leads to insufficient embedding of dynamics within a feasibly small number of modes. To take into account the effects of the discarded modes, several closure modeling approaches are devised (see for instance \cite{couplet2003intermodal,kalb2007intrinsic,bergmann2009enablers,wang2012proper,baiges2015reduced,xie2017approximate}), which serve a dual purpose: that of numerical stabilization as well as statistical fidelity preservation. Following the large eddy simulation (LES) ideas, it has been shown that the eddy viscosity concept provides an efficient framework to account the effect of the truncated modes \cite{akhtar2012new,san2014proper,san2015stabilized}. In this study, we put forth a robust dynamic procedure for computing the modal eddy viscosities in order to stabilize the ROMs. The novelty of our approach stems from the design of an artificial neural network (ANN) architecture to predict the magnitude of the mode dependent eddy viscosity dynamically, thus removing the need for an a-priori specification of an arbitrary value.

ANNs and other machine learning (ML) strategies have engendered a revolution in data-driven prediction applications and are seeing widespread investigation in the computational physics community. Previous studies into the feasibility of similar ML techniques for ROMs of various nonlinear systems may be found in \cite{narayanan1999low,sahan1997artificial,moosavi2015efficient,san2018neural}.
In particular, we have recently illustrated the ANN concept for model order reduction of the one-dimensional Burgers equation and the performance of several training algorithms has been documented \cite{san2018neural}. In the present study, however, we put forth a modified ANN architecture since it is more appropriate to turbulent flows. ML approaches have also been developed for use in feedback flow control where they generate a direct mapping of flow measurements to actuator control systems \cite{gillies1998low,faller1997unsteady,efe2004modeling,lee1997application}. In our investigation, information from the high fidelity evolution of governing laws is leveraged to provide a supervised learning framework for a single layer ANN to stabilize ROMs of the mesoscale forced-dissipative geophysical turbulence system. In brief, an ANN estimates a nonlinear relationship between a desired set of inputs and targets provided viable benchmark data for their underlying statistical relationship is available. This subset of the ML field has seen wide application in function approximation, data classification, pattern recognition and dynamic systems control applications \cite{widrow1994neural,demuth2014neural} and is generating great interest for its utility in the reproduction of systems with pronounced nonlinear interactions \cite{raissi2017physics1}.

Before its deployment as a prediction or regression tool, an ANN is trained to accurately capture the nonlinear relationship between its inputs and outputs through some classical loss function (such as mean squared error). A \emph{regularized} training ensures that the framework avoids overfitting any noise that may have been present in the training data. For our supervised learning framework, we utilize the extreme learning machine (ELM) \cite{huang2006extreme} training procedure, which stands out from other machine learning methods with a direct (i.e., non-iterative) fast training capability. ELM is a kind of regularized neural network where the weights connecting inputs and hidden nodes are randomly assigned and never updated. The output weights of hidden nodes are then learned in a single step using a pseudoinverse approach, which provides an extremely fast learning mechanism, in the least squares sense, compared to the networks trained using traditional backpropagation approaches \cite{cancelliere2017analysis}. For our investigation it is seen that a single hidden layer feed-forward neural networks (SLFN) ELM algorithm satisfies generalized training requirements with extremely reduced computational cost yet substantially accurate reproductions of training statistics.

For assessing our proposed framework, we utilize the governing laws given by the barotropic vorticity equation (BVE) model. It is a simplified two-dimensional framework, also known as the one-layer quasigeostrophic (QG) model \cite{majda2006nonlinear}, commonly used to study mesoscale ocean circulation problems. While the POD model reduction framework has been used to derive ROMs of the BVE (see, e.g., \cite{selten1995efficient,crommelin2004strategies}), the present work represents an attempt to model the unrepresented scales of the QG dynamics, mesoscale turbulence and their effect on mean circulation using an ANN based supervised machine learning framework. The novelty of our approach is therefore adding modal dissipation that is correlated to modal amplitude via a neural net. Our method can be considered as a hybrid modeling paradigm combining machine learning principles and physics-based simulation tools for QG dynamics. The decision to choice the ELM training approach is to ensure robust generalization for such noisy data.


\section{Full Order Modeling}
Oceanic and atmospheric flows display an enormous range of spatial and temporal scales, from seconds to decades and from centimeters to thousands of kilometers. Thus, a model incorporating all the relevant physics of the ocean and atmosphere would be impractical for numerical simulations. During the last decades, significant advancements were made in developing simplified models for geophysical fluid dynamics \cite{mcwilliams2006fundamentals}, which have been instrumental in providing relatively accurate numerical results at a reasonable computational price. Although these models have continued to produce increasingly accurate results and therefore improved weather forecasting, their use in long time integrations such as those required by climate modeling remains challenging \cite{ghil2008climate,lynch2008origins}. To illustrate our surrogote proposed framework, we consider the BVE model, which has been extensively used to study forced-dissipative QG dynamics \cite{majda2006nonlinear}. The dimensionless BVE may be given by \cite{greatbatch2000four,san2011approximate}
\begin{equation}\label{eq:nbve}
\frac{\partial \omega}{\partial t} + J(\omega,\psi) -\frac{1}{Ro}\frac{\partial \psi}{\partial x} = \frac{1}{Re}\nabla^2 \omega + \frac{1}{Ro}\sin(\pi y),
\end{equation}
where $\omega$ is the kinematic vorticity and $\psi$ is the streamfunction. The nonlinear advection term is defined by the Jacobian
\begin{equation}\label{eq:jac}
J(\omega,\psi) =  \frac{\partial \psi}{\partial y}\frac{\partial \omega}{\partial x} - \frac{\partial \psi}{\partial x}\frac{\partial \omega}{\partial y},
\end{equation}
since we define the flow velocity components by
\begin{equation}\label{eq:vel}
u = \frac{\partial \psi}{\partial y}, \quad v =- \frac{\partial \psi}{\partial x},
\end{equation}
and the following kinematic relationship holds for satisfying the incompressibility constraint
\begin{equation}\label{eq:poi}
\nabla^2 \psi = -\omega,
\end{equation}
where $\nabla^2$ is the standard Laplacian operator. The dimensionless BVE given in Eq.~(\ref{eq:nbve}), has two nondimensional parameters, the Reynolds and Rossby numbers, which are related to the characteristic length and velocity scales in the following way:
\begin{equation}\label{eq:ReRo1}
Re = \frac{V L}{\nu}, \quad Ro = \frac{V}{\beta L^2}.
\end{equation}
where $\nu$ is the horizontal eddy viscosity of the BVE model and $\beta$ is the Rossby parameter. We note that Eq.~(\ref{eq:nbve}) uses the $\beta$-plane approximation, valid for most large-scale ocean basins, which accounts for the Earth's rotational effects by approximating the Coriolis parameter. For the purpose of nondimensionalization, $L$ represents a characteristic horizontal length scale given by the basin dimension in the $x$ direction, and $V$ is a reference velocity scale (also known as the Sverdrup velocity) given by
\begin{equation}\label{eq:sverdrup}
V = \frac{\tau_0}{\rho H}\frac{\pi}{\beta L},
\end{equation}
where $\tau_0$ is the maximum amplitude of the sinusoidal double-gyre wind stress, $\rho$ is the reference fluid density, and $H$ is the reference depth of the ocean basin. Following \cite{greatbatch2000four,nadiga2001dispersive,san2011approximate}, we consider a four-gyre circulation problem, a benchmark oceanic flow problem whose behavior is difficult to capture correctly in coarse grained models \cite{san2011approximate}. Indeed, as shown in \cite{san2015stabilized}, the standard model order reduction approaches without stabilization are incapable of resolving the correct physics. In our full order model (FOM) simulations we use a second-order accurate kinetic energy and enstrophy conserving Arakawa finite difference scheme \cite{arakawa1966computational}. The derivatives in the linear terms are also approximated using the standard second-order finite differences. Our time advancement scheme is given by the classical total variation diminishing third-order accurate Runge-Kutta scheme. Details of the Poisson solver, numerical schemes and boundary conditions used for this study may be found in \cite{san2011approximate}.

\section{Reduced order modeling}
\vspace{-2pt}
\label{sec:rom}
We build our reduced order modeling framework based on a standard projection methodology using the method of snapshots \cite{sirovich1987turbulence}. Solving the FOM given by Eq.~(\ref{eq:nbve}), the $n$th record of the prognostic variable (vorticity field) is denoted $\omega(x,y,t_n)$ for $n=1,2, ..., N$, where $N$ is the number of snapshots recorded for basis construction. Then we decompose
the solution field into a time invariant averaged $\bar{\omega}(x,y)$ and a fluctuating component $\omega'(x,y,t)$ through \cite{holmes1998turbulence,noack2003hierarchy},
\begin{align}
    \label{mat_rev_1}
    \omega(x,y,t) = \bar{\omega}(x,y) + \omega'(x,y,t) \quad x,y \in \Omega \, ,
\end{align}
where $\Omega$ is the two-dimensional domain and the mean of the snapshot data is
\begin{align}
    \label{mat_rev_44}
    \bar{\omega}(x,y) = \frac{1}{N} \sum_{n=1}^{N} \omega(x,y,t_n).
\end{align}
In order to obtain the POD basis functions, a correlation matrix of the fluctuating part is constructed by
\begin{equation}\label{eq:cor}
    a_{ij}=\int_{\Omega} \omega'(x,y,t_i) \omega'(x,y,t_j) dx dy,
\end{equation}
where the subscripts $i$ and $j$ refer to snapshot indexes. We must note that the data correlation matrix $\mathbf{A}=[a_{ij}]$ is a non-negative Hermitian matrix. We further define the inner product of two functions $f$ and $g$ as
\begin{equation}\label{eq:inner}
    \langle f,g\rangle=\int_{\Omega} f g dx dy,
\end{equation}
such that Eq.~(\ref{eq:cor}) yields $a_{ij}=\langle\omega'(x,y,t_i), \omega'(x,y,t_j)\rangle$. The optimal POD basis functions may then be obtained by solving the following eigenvalue problem \cite{ravindran2000reduced}
\begin{equation}\label{eq:eig}
    \mathbf{A} \mathbf{\Gamma} =\mathbf{\Gamma} \mathbf{\Lambda},
\end{equation}
where $\mathbf{\Lambda}=\mbox{diag}[\lambda_1, ..., \lambda_N ]$ is the diagonal eigenvalue matrix and $\mathbf{\Gamma}=[\boldsymbol\gamma_1, ...,\boldsymbol\gamma_N]$ refers to right eigenvector matrix whose columns are eigenvectors of the correlation matrix $\mathbf{A}$. The eigenvalues are usually stored in descending order for practical purposes i.e., $\lambda_1\geq\lambda_2\geq ...\geq \lambda_N$. Then the orthogonal POD basis functions of the vorticity field can be obtained as
\begin{equation}\label{eq:PODbasis}
    \phi_{k}(x,y) =\frac{1}{\sqrt{\lambda_{k}}} \sum_{n=1}^{N} \gamma_{nk}\omega'(x,y,t_n),
\end{equation}
where $\lambda_{k}$ is the $k$th eigenvalue, $\gamma_{nk}$ is the $n$th component of the $k$th eigenvector, and $\phi_{k}(x,y)$ is the $k$th POD mode. The dependence between streamfunction and vorticity given by Eq.~(\ref{eq:poi}) may be utilized to obtain the $k$th basis function for the streamfunction, $\varphi_{k}(x,y)$, by solving a Poisson equation
\begin{equation}\label{eq:psi}
    \nabla^2 \varphi_{k} = -\phi_{k}.
\end{equation}

Now we can span our field variables into the POD modes as follows
\begin{equation}\label{eq:const1}
    \omega(x,y,t) = \bar{\omega}(x,y) + \sum_{k=1}^{M} \alpha_{k}(t)\phi_{k}(x,y),
\end{equation}
\begin{equation}\label{eq:const1}
    \psi(x,y,t) = \bar{\psi}(x,y) + \sum_{k=1}^{M} \alpha_{k}(t)\varphi_{k}(x,y),
\end{equation}
where we have decomposed $\omega'(x,y,t)$ using time dependent modal coefficient $\alpha_k$ and the POD modes $\phi_{k}(x,y)$. We note that the kinematic relationship given by Eq.~(\ref{eq:psi}) implies that the same $\alpha_k$ accounts for the streamfunction as well.
A ROM can be generated by a truncation of the $N$ total bases to only $M$ retained modes where $M\ll N$. These largest energy containing modes correspond to the $M$ largest eigenvalues, $\lambda_1$, $\lambda_2$, ..., $\lambda_M$. To obtain our standard ROM, an orthogonal Galerkin projection is performed by multiplying Eq.~(\ref{eq:nbve}) with the POD basis functions and integrating over the domain $\Omega$. The resulting dynamical system for $\alpha_k$ can be written as
\begin{equation} \label{eq:rom1}
  \frac{d \alpha_k}{d t} = \mathfrak{B}_{k} + \sum_{i=1}^{M} \mathfrak{L}^{i}_{k}\alpha_{i} + \sum_{i=1}^{M}\sum_{j=1}^{M} \mathfrak{N}^{ij}_{k}\alpha_{i}\alpha_{j},
\end{equation}
where
\begin{eqnarray}
  & & \mathfrak{B}_{k} = \big\langle \frac{1}{Re}\nabla^2 \bar{\omega} + \frac{1}{Ro}\sin(\pi y) +\frac{1}{Ro}\frac{\partial \bar{\psi}}{\partial x} - J(\bar{\omega},\bar{\psi}), \phi_{k} \big\rangle , \nonumber \\
  & & \mathfrak{L}^{i}_{k} = \big\langle \frac{1}{Re}\nabla^2 \phi_{i} +  \frac{1}{Ro}\frac{\partial \varphi_{i}}{\partial x} - J(\bar{\omega},\varphi_{i}) -  J(\phi_{i},\bar{\psi}) , \phi_{k} \big\rangle  , \nonumber\\
  & &  \mathfrak{N}^{ij}_{k} =  \big\langle -J(\phi_{i},\varphi_{j}), \phi_{k} \big\rangle. \label{eq:roma7}
\end{eqnarray}
The ROM given by Eq.~(\ref{eq:rom1}) consists of $M$ coupled ordinary differential equations (ODEs) for modal coefficients, which are solved numerically by a third-order Runge-Kutta scheme. We note that the resulting ROM is highly efficient since both the POD basis functions and the coefficients of ODEs given by Eq.~(\ref{eq:roma7}) can be precomputed from the data provided by snapshots. A complete specification of the dynamical system given by Eq.~(\ref{eq:rom1}) may be obtained by the following projection of the initial condition:
\begin{equation}\label{eq:ic1}
\alpha_{k}(t_{0}) = \big\langle \omega(x,y,t_{0}) - \bar{\omega}(x,y), \phi_{k} \big\rangle,
\end{equation}
where $\omega(x,y,t_{0})$ is the vorticity field specified at time $t_{0}$.

The standard ROM given by Eq.~(\ref{eq:rom1}) usually works well for a periodic or quasi-periodic system for which the first few POD modes can capture the system's dynamics. However, one of the main sources of inaccuracy in a truncated ROM framework is the potential for instability due to neglecting the contributions of the higher POD modes. Therefore, many stabilization schemes are utilized in order to improve the performance of the ROMs \cite{couplet2003intermodal,kalb2007intrinsic,bergmann2009enablers,wang2012proper,baiges2015reduced,xie2017approximate}).
Using an eddy viscosity approach, the stabilization of the ROM can be achieved by \cite{san2014proper,san2015stabilized}
\begin{equation} \label{eq:rom2}
  \frac{d \alpha_k}{d t} = \mathfrak{B}_{k} +\tilde{\mathfrak{B}}_{k}+ \sum_{i=1}^{M} (\mathfrak{L}^{i}_{k}+\tilde{\mathfrak{L}}^{i}_{k})\alpha_{i} + \sum_{i=1}^{M}\sum_{j=1}^{M} \mathfrak{N}^{ij}_{k}\alpha_{i}\alpha_{j},
\end{equation}
where, using the Smagorinsky model and the analogy between ROM and LES, two additional terms can be written as
\begin{align}\label{EVstab4}
  \tilde{\mathfrak{B}}_{k} &= \langle \nu^{e}_{k} \nabla^2 \bar{\omega},\phi_k\rangle , \nonumber\\
  \tilde{\mathfrak{L}}^{i}_{k} &= \langle\nu^{e}_{k}  \nabla^2 \phi_i, \phi_k\rangle ,
\end{align}
where $\nu^{e}_{k}$ is the modal eddy viscosity parameter. This free stabilization parameter may be simply considered as a global constant for all the modes \cite{aubry1988dynamics,wang2012proper}. The global constant eddy viscosity idea may improved by supposing that the amount of dissipation is not identical for all the POD modes. It has been shown that finding an optimal value for this parameter significantly improves the predictive performance of ROMs \cite{san2014proper,san2015stabilized}. Therefore, the chief novelty of the present study is the utilization of a novel ML framework to estimate these modal eddy viscosity coefficients to stabilize and overcome errors due to the finite truncation in ROMs. We determine $\nu^{e}_{k}$ dynamically from our ML framework during the evolution of each temporal mode $\alpha_k$ at each time step.

\section{Artificial Neural Network Architecture}
\label{sec:ML}
In this section, we introduce an SLFN architecture for predicting modal eddy viscosity coefficients for stabilization of ROMs. Figure \ref{fig:nn_schem} illustrates our ANN architecture which consists of an input layer, a hidden layer and an output layer. Each layer possesses a predefined number of nodes called neurons. Except for the input neurons, each neuron has an associated bias and activation function. The main goal in any supervised learning framework is to find a mapping between input nodes and output nodes. Mathematically, we are looking for a mapping $\mathfrak{M}$ to establish a relationship between input nodes $x_p$ and output nodes $y_j$ as follows:
\begin{figure}[htbp]
\centering
\includegraphics[width=\mywe]{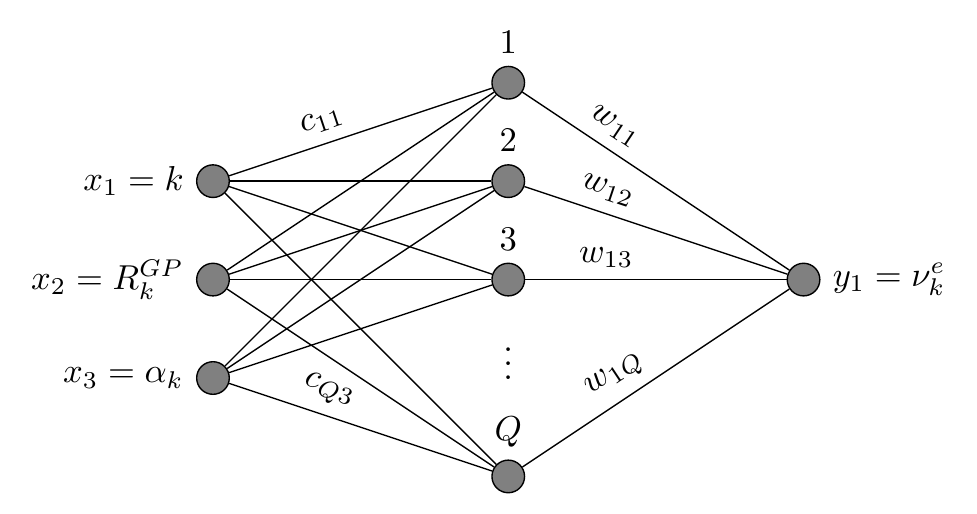}
\caption{A schematic of the SLFN utilized for the stabilization framework in this study. Our inputs are resolved ROM variables ($P=3$) and our prediction is a mode dependent eddy-viscosity ($J=1$).}
\label{fig:nn_schem}
\end{figure}
\begin{align}\label{ANN_2R1}
\mathfrak{M}: \{x_1,x_2,...,x_P\} \in \R^P \rightarrow \{y_1,y_2,...,y_J\} \in \R^J,
\end{align}
where $P$ is the number of input neurons and $J$ is the number of output neurons. If $Q$ refers to the number of hidden layer neurons, the $j$th output node can be computed as
\begin{align}\label{pred}
y_j = G\Big(d_j + \sum_{q=1}^{Q} w_{jq} F\big(b_{q} + \sum_{p=1}^{P} c_{qp} x_{p} \big)\Big)
\end{align}
where $c_{qp} \in \R^{Q \times P}$ are the connection weights between the neurons in input and hidden layers, and $w_{jq} \in \R^{J \times Q}$ are the weights between the neurons in hidden and output layers. Here, $F$ and $G$ are neurons' activation functions; and $b_q \in \R^{Q}$ and $d_j \in \R^{J}$ are called biases operating as thresholds for hidden and output layers, respectively. In this study, we have utilized the tan-sigmoid activation function for the hidden layer neurons, which can be expressed as
\begin{align}\label{ANN_3}
  F(s) = \frac{2}{1+\exp(-2s)}-1,
\end{align}
and a linear activation function for the output layer neurons given by
\begin{align}\label{ANN_3a}
  G(s) = s.
\end{align}
While it has been reported that sigmoidal activation functions saturate across a large portion of their domain \cite{goodfellow2016deep}, our reasoning behind the use of the classical tan-sigmoid activation was to leverage the benefit of the saturation behavior to obtain a bounded prediction from the network.

\subsection{Extreme learning machine}
\label{subsec:ELM}
Introducing $N$ sample training data examples (i.e., input-output pairs), the weights and biases can be computed in a supervised learning framework using either well established iterative back propagation methods \cite{carrillo2016estimation} or pseudoinverse approaches \cite{cancelliere2017analysis}. As mentioned previously, the ANN architecture is trained by utilizing an ELM approach proposed in \cite{huang2006extreme} for extremely fast training of an SLFN. The ELM approach requires no biases in the output layer (i.e., $d_j=0$). In the ELM method, the weights $c_{qp}$ and biases $b_{q}$ are initialized randomly from a uniform distribution (i.e., between -1 and 1 in our study) and no longer modified. Therefore the only unknowns to be determined are $w_{jq}$ weights. Using the linear activation function for the output layer, Eq.~(\ref{pred}) can be written for $N$ sample examples
\begin{align}\label{pred2}
y_{jn} =  \sum_{q=1}^{Q} w_{jq} F\big(b_{q} + \sum_{p=1}^{P} c_{qp} x_{pn} \big)
\end{align}
where $x_{pn} \in \R^{P \times N}$ and  $y_{jn} \in \R^{J \times N}$ refer to the training input-output data pairs. Using a more convenient matrix notation (i.e., $\mathbf{X} =[x_{pn}]$, $\mathbf{Y}=[y_{jn}]$, $\mathbf{C}=[c_{qp}]$, $\mathbf{W} =[w_{jq}]$, and $\mathbf{b}=[b_{q}]$), our learning problem can be written as
\begin{align}\label{pred3}
\mathbf{Y} =  \mathbf{W} \mathbf{H}^{\intercal}
\end{align}
where $\mathbf{H}^{\intercal} \in \R^{Q \times N}$ is given by
\begin{align}\label{pred4}
\mathbf{H}^{\intercal} = F(\mathbf{B} + \mathbf{C} \mathbf{X})
\end{align}
where the vector $\mathbf{b}$ is repeated across $N$ columns as shown in Eq.~(\ref{pred2}) (i.e., $\mathbf{B}=[\mathbf{b}, \mathbf{b}, ..., \mathbf{b}] \in \R^{Q \times N}$).
By taking the transpose of both side of Eq.~(\ref{pred3}) we can write
\begin{align}\label{pred5}
\mathbf{H}\mathbf{W}^{\intercal} = \mathbf{Y}^{\intercal}
\end{align}
and the solution for the weights can be computed by
\begin{align}\label{pred6}
\mathbf{W}^{\intercal} = \mathbf{H}^{\dagger} \mathbf{Y}^{\intercal}
\end{align}
where $\mathbf{H}^{\dagger} \in \R^{Q \times N}$ is the pseudoinverse of $\mathbf{H} \in \R^{N \times Q}$. In order to compute the pseudoinverse, we apply the following singular value decomposition (SVD) to the matrix $\mathbf{H}$ since its number of rows is greater than its number of columns in typical ML applications (i.e., $N \geq Q$)
\begin{align}\label{pred7}
\mathbf{H} = \mathbf{U} \mathbf{\Sigma} \mathbf{V}^{\intercal}
\end{align}
where $\mathbf{U} \in \R^{N \times Q}$ and $\mathbf{V} \in \R^{Q \times Q}$ are column-orthogonal and orthogonal matrices, and $\mathbf{\Sigma}\in \R^{Q \times Q}$ is a diagonal matrix whose elements (i.e., $\sigma_{qq}=\sigma_q$) are non-negative and called singular values. Using the SVD, the pseudoinverse of $\mathbf{H}$ becomes
\begin{align}\label{pred8}
\mathbf{H}^{\dagger} = \mathbf{V} \mathbf{\Sigma}^{\dagger} \mathbf{U}^{\intercal}
\end{align}
where $\mathbf{\Sigma}^{\dagger}$ can be computed from $\mathbf{\Sigma}$ by taking the reciprocal of each non-zero element (i.e., $\sigma_{q}^{\dagger}=1/\sigma_{q}$). However, the presence of tiny singular values can cause numerical instability. Therefore, a well-known Tikhonov type regularization is often introduced by
\begin{align}\label{pred8}
\sigma_{q}^{\dagger} = \frac{\sigma_{q}}{\sigma_{q}^{2} + \epsilon}
\end{align}
where $\lambda$ controls the trade off between the least-squares error and the penalty term for regularization (e.g., see  \cite{cancelliere2017analysis}). In the present study we set $\epsilon=10^{-12}$. Finally, using Eq.~(\ref{pred6}), the unknown weights can be computed by
\begin{align}\label{pred9}
\mathbf{W} = \mathbf{Y} \mathbf{U} \mathbf{\Sigma}^{\dagger} \mathbf{V}^{\intercal}.
\end{align}

\subsection{Training data}
\label{subsec:Training}

Our architecture is devised to take inputs accessible to us during the time integration of the ROM and estimate the modal eddy viscosity coefficient. Our high fidelity snapshot data (from which POD bases are constructed) are also used to train our architecture. First, we denote the right hand side of Eq.~(\ref{eq:rom1}) as
\begin{align}\label{eq:Training1}
  R_k^{GP} = \mathfrak{B}_{k} + \sum_{i=1}^{M} \mathfrak{L}^{i}_{k}\alpha_{i} + \sum_{i=1}^{M}\sum_{j=1}^{M} \mathfrak{N}^{ij}_{k}\alpha_{i}\alpha_{j},
\end{align}
and then apply the Galerkin projection to FOM given by Eq.~(\ref{eq:nbve}), which yields the true solution
\begin{align}\label{eq:Training2}
  R_k^{FOM} = \Big\langle  \frac{1}{Re}\nabla^2 \omega + \frac{1}{Ro}(\sin(\pi y) +\frac{\partial \psi}{\partial x}) -J(\omega,\psi),\phi_{k} \Big\rangle.
\end{align}
The ideal stabilization would thus conform to the differences between these quantities i.e.,
\begin{align}\label{eq:Training3}
  \tilde{R}_k = R_k^{FOM} - R_k^{GP}.
\end{align}
We know from Eq.~(\ref{EVstab4}) that
\begin{align}\label{eq:Training4}
  \tilde{R}_k = \nu^{e}_{k} \Big( \langle \nabla^2 \bar{\omega},\phi_k\rangle + \sum_{i=1}^{M}\langle\nabla^2 \phi_i, \phi_k\rangle \alpha_i \Big),
\end{align}
where we redefine
\begin{align}\label{eq:Training5}
  R_{k}^{STAB} = \langle \nabla^2 \bar{\omega},\phi_k\rangle + \sum_{i=1}^{M}\langle\nabla^2 \phi_i, \phi_k\rangle \alpha_i ,
\end{align}
and therefore we compare Eq.~(\ref{eq:Training3}) and Eq.~(\ref{eq:Training4}) to obtain the modal eddy viscosity coefficients
\begin{align}\label{eq:Training6}
  \nu^{e}_{k} = \frac{R_k^{FOM} - R_k^{GP}}{R_k^{STAB}},
\end{align}
as the eddy viscosity stabilization for each mode within the training data set. Although Eq.~(\ref{eq:Training6}) is an exact relationship, we use a clipping procedure for numerical stability by discarding negative entries of $\nu^{e}_{k}$ in our training data set. Therefore, our training data is generated by considering the following bounds
\begin{align}\label{eq:Training88}
  \epsilon = 10^{-12}\leq \nu^{e}_{k} \leq \frac{c}{Re}=c\frac{\nu}{VL},
\end{align}
where $c$ is the upper bound of the relative ratio between the stabilized viscosity and physical model viscosity. In the present study, we set $c=6$, which provides six times larger stabilization viscosity than the specified $\nu$ of the original model. We have also verified that the proposed ROM-ANN approach is robust to the selection of $c$ (i.e., similar statistical results have been observed for $c=4$ and $c=10$ sets). The clipping approach presented by Eq.~(\ref{eq:Training88}) can be considered as a physical realizability bounds of ROM training data. With this realizable calculation of the stabilization viscosity, we hypothesize that a mode dependent nonlinear (but unknown) relationship exists between the resolved modes in the ROM that estimates $\nu^{e}_{k}$ dynamically. To conclude, our ANN framework is trained between inputs given by the modal index $k$, $R_k^{GP}$, and $\alpha_k$ (i.e., they are all available during the ROM time stepping) and to predict an approximation for $\nu^{e}_{k}$. We thus have 3 inputs to our network with $Q$ hidden layer neurons to obtain 1 output (which is the modal eddy viscosity coefficient). The architecture of our ANN is shown in Figure \ref{fig:nn_schem}. We emphasize that this simplified ANN is basically a non-linear regression or a curve fitting between input and target states. As we will show in next section, however, its generalization is quite remarkable for both in-sample data and out-of-sample data predictions.  

\section{Results}
\label{sec:results}

To validate our proposed ANN framework, we consider the four-gyre barotropic circulation problem \cite{greatbatch2000four,nadiga2001dispersive,san2011approximate}).  This test problem yields four gyres circulation patterns in the time mean in a shallow ocean basin and represents an ideal test for the viability of the proposed ROM. Indeed it was shown that ROMs without stabilization are incapable of resolving the mean dynamics \cite{san2015stabilized}.

The dimensionless form of the BVE describing the QG problem is evolved from $t=0$ to $t=100$ using a fixed time step $\Delta t = 2.5 \times 10^{-5}$ on a Munk layer resolving $256 \times 512$ computational grid resolution. The dimensionless parameters of the BVE system are chosen as $Re=450$ and $Ro=3.6 \times 10^{-3}$. We must note that $t=10$ to $t=100$ is our data collection window (for the purpose of POD basis generation as well as ANN training) due to a statistically steady state reached after the initial transient period. 900 snapshots are collected during this period which are equally distributed in time. The ideal eddy viscosity is also computed at these snapshots for training our machine learning framework. We note here that our ROM (whether purely truncated or stabilized by ANN) is utilized for predictions upto $t=200$ utilizing the POD modes obtained from our previously mentioned data collection window. This may be considered to be a challenging validation of our dual data-driven methodology for the QG problem. For both our standard ROM and stabilized ROM-ANN computations, the same time step with $\Delta t = 2.5 \times 10^{-5}$ is used for time integration of the dynamical system. Sensitivity studies for varying time steps will be also presented later.

\begin{table}[]
\centering
\label{tab:1}
\caption{L\textsubscript{2}-norm errors of the reduced order models (with respect to FOM) for the mean vorticity and streamfunction fields. Note that the ROM-ANN retains only $M=10$ modes.}
\begin{tabular}{p{0.20\textwidth}p{0.12\textwidth}p{0.12\textwidth}}
\hline\noalign{\smallskip}
& Vorticity & Streamfunction \\
\noalign{\smallskip}\hline\noalign{\smallskip}
\multicolumn{2}{l}{\textsl{\underline{No stabilization}}} \\
ROM ($M=10$)  & $6.89 \times 10^7$ & $1.50 \times 10^5$  \\
ROM ($M=20$) & $2.85 \times 10^5$ & $5.69 \times 10^2$  \\
ROM ($M=30$) & $1.07 \times 10^3$ & $1.14 \times 10^0$  \\
\multicolumn{2}{l}{\textsl{\underline{With stabilization}}} \\
ROM-ANN ($Q=20$)  & $1.20 \times 10^3$ & $9.45\times 10^{-1}$  \\
ROM-ANN ($Q=40$) & $6.54 \times 10^2$ & $1.22\times 10^{-1}$ \\
ROM-ANN ($Q=60$) & $4.31 \times 10^2$ & $3.09 \times 10^{-1}$  \\
\noalign{\smallskip}\hline
\end{tabular}
\end{table}

\begin{figure}[htbp]
\centering
\mbox{
\subfigure{\includegraphics[width=\mywd]{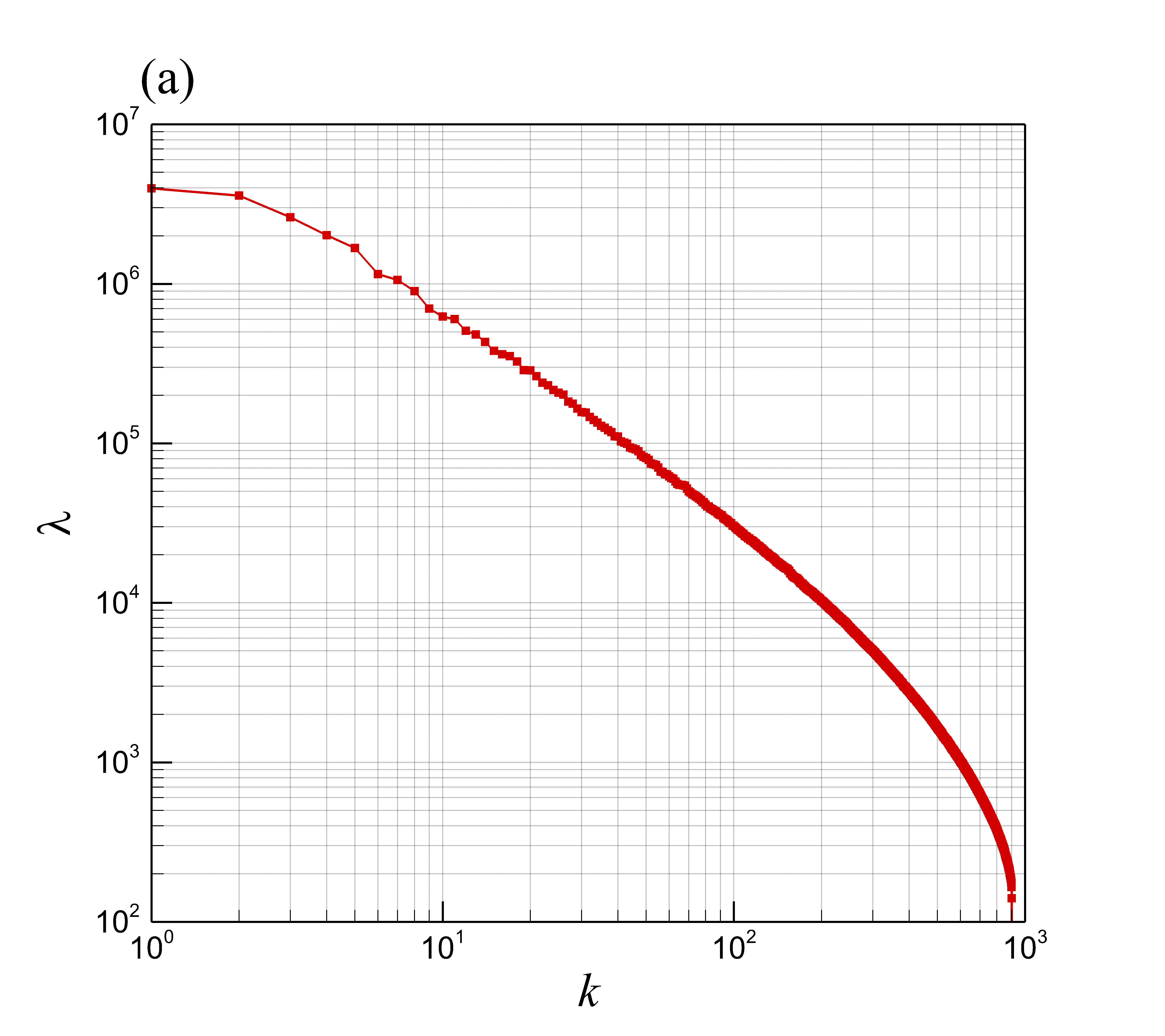}}
}\\
\mbox{
\subfigure{\includegraphics[width=\mywd]{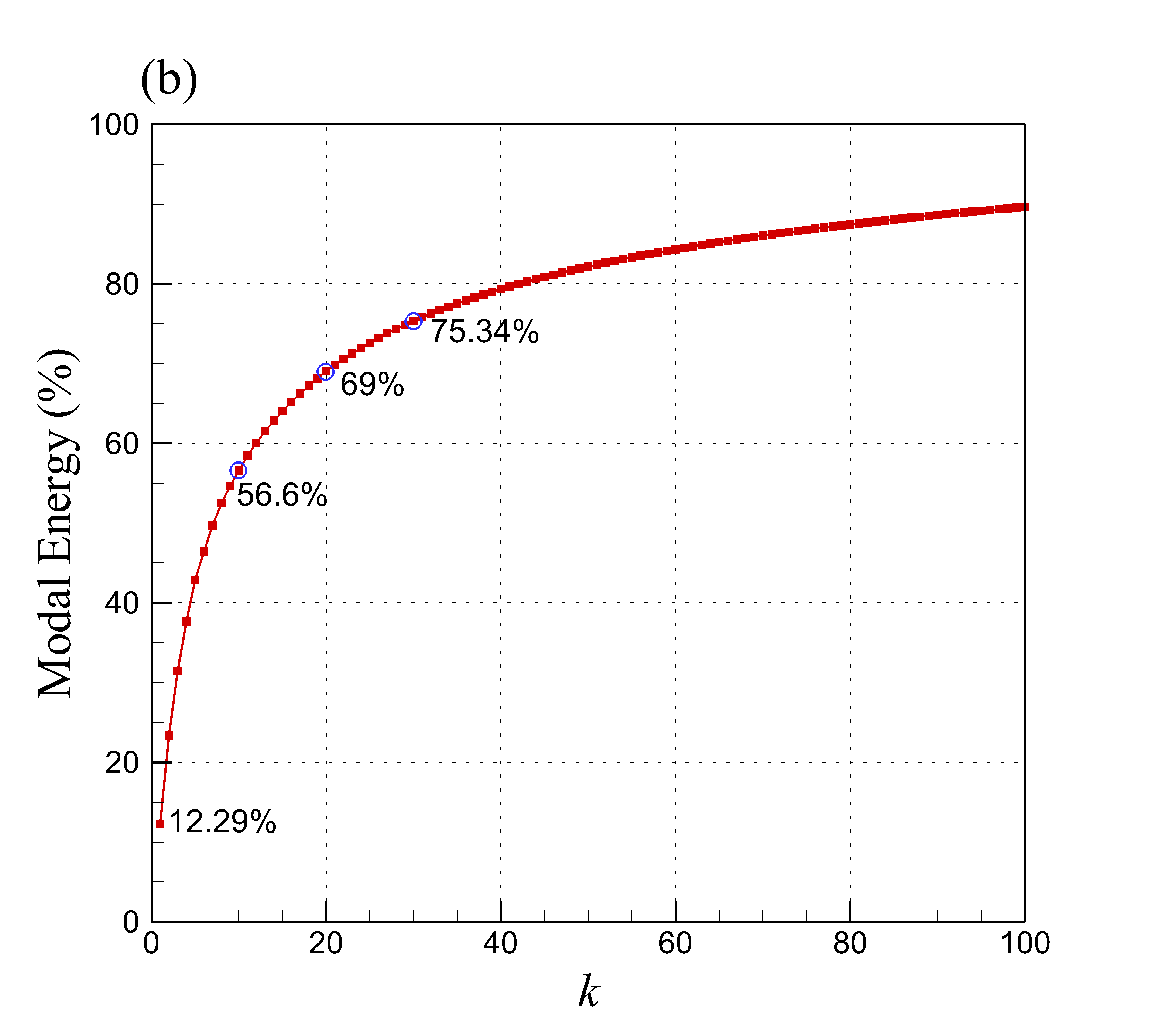}}
}
\caption{POD analysis by the snapshot data for $Re=450$ and $Ro=3.6 \times 10^{-3}$.  (a) The distribution of eigenvalues; and (b) their energy levels with $k$ denoting the modal index.}
\label{fig:u1}
\end{figure}

\begin{figure}[htbp]
\centering
\mbox{
\subfigure{\includegraphics[width=\mywa]{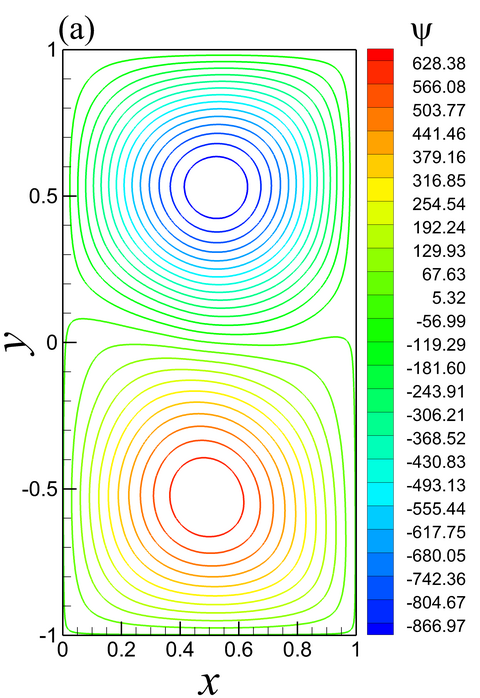}}
\subfigure{\includegraphics[width=\mywa]{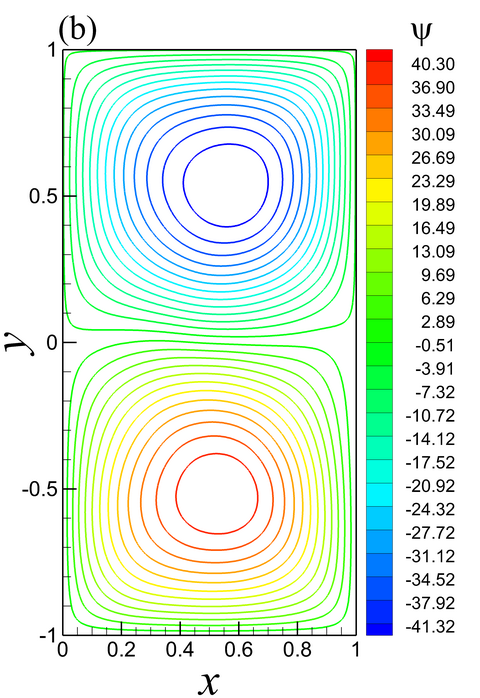}}
\subfigure{\includegraphics[width=\mywa]{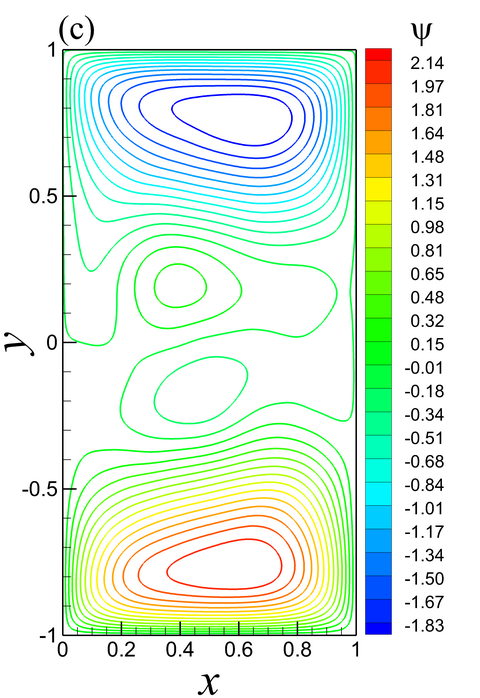}}
}\\
\mbox{
\subfigure{\includegraphics[width=\mywa]{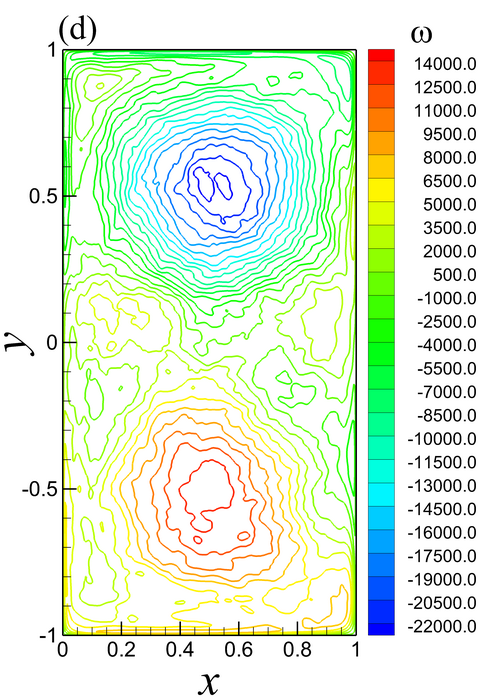}}
\subfigure{\includegraphics[width=\mywa]{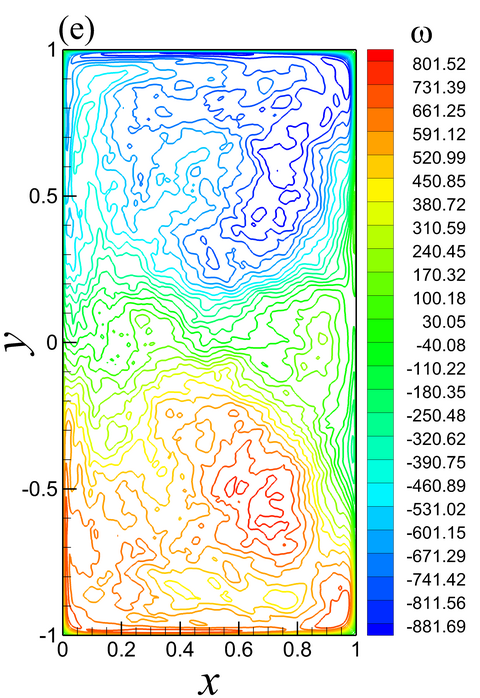}}
\subfigure{\includegraphics[width=\mywa]{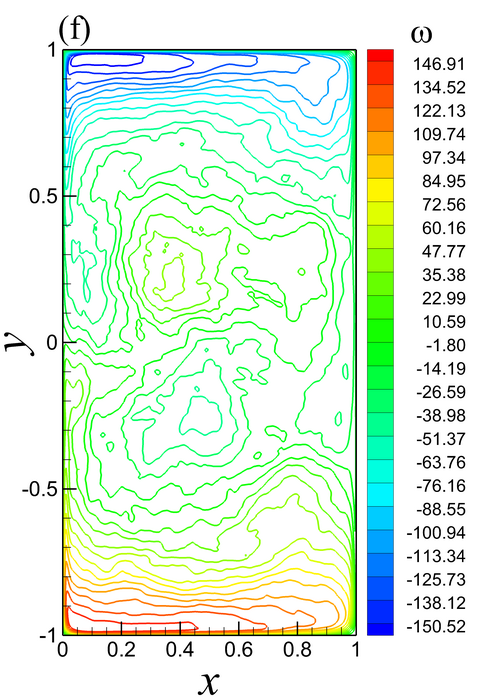}}
}
\caption{Mean streamfunction and vorticity fields retrieved from the standard Galerkin projection method. (a) $\psi$ with $M=10$ modes; (b) $\psi$ with $M=20$ modes; (c) $\psi$ with $M=30$ modes; (d) $\omega$ with $M=10$ modes; (e) $\omega$ with $M=20$ modes; and (f) $\omega$ with $M=30$ modes.}
\label{fig:u4}
\end{figure}

Figure \ref{fig:u1} shows the accumulation of energies in the form of eigenvalue magnitudes where it can be seen that a large majority (close to 75\%) of the energies are accumulated in the first 30 modes of the transformed space. Figure \ref{fig:u4} shows the gradual convergence of the ROM (i.e., without stabilization) to the four-gyre circulation pattern with increasing $M$. Indeed, non-physical two-gyre pattern is observed for the case of $M=10$ and $M=20$.


%

\begin{figure}[htbp]
\centering
\mbox{
\subfigure{\includegraphics[width=\mywa]{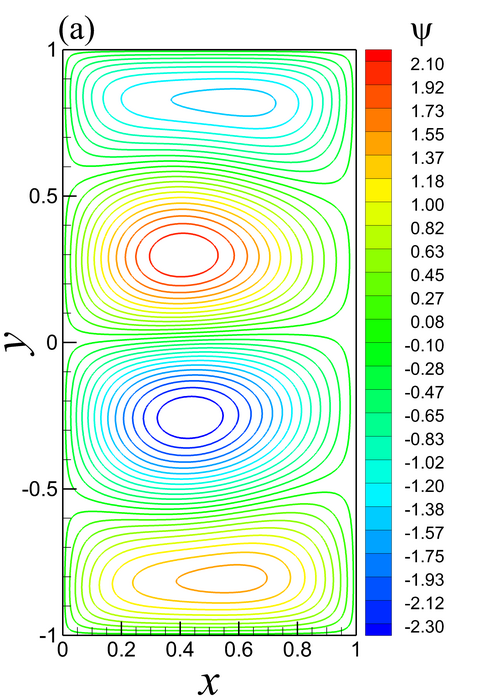}}
\subfigure{\includegraphics[width=\mywa]{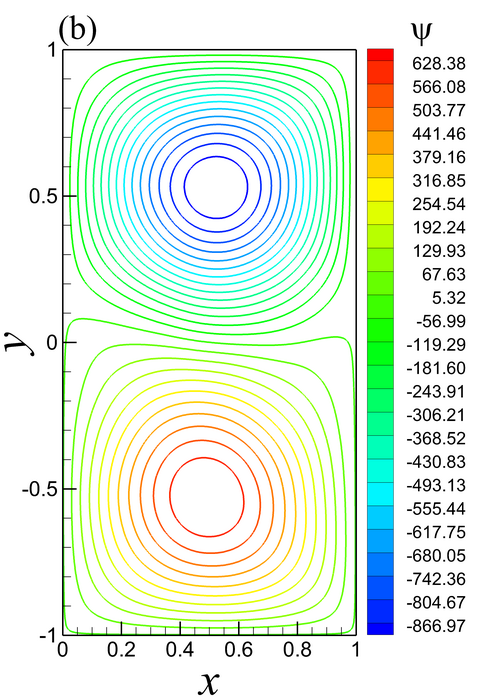}}
\subfigure{\includegraphics[width=\mywa]{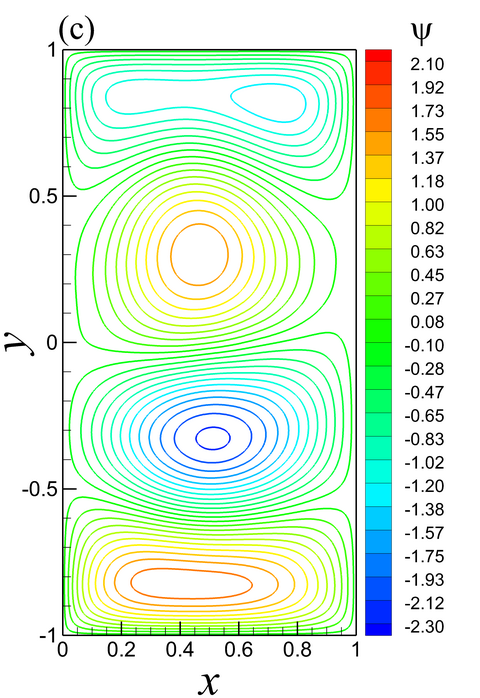}}
}\\
\mbox{
\subfigure{\includegraphics[width=\mywa]{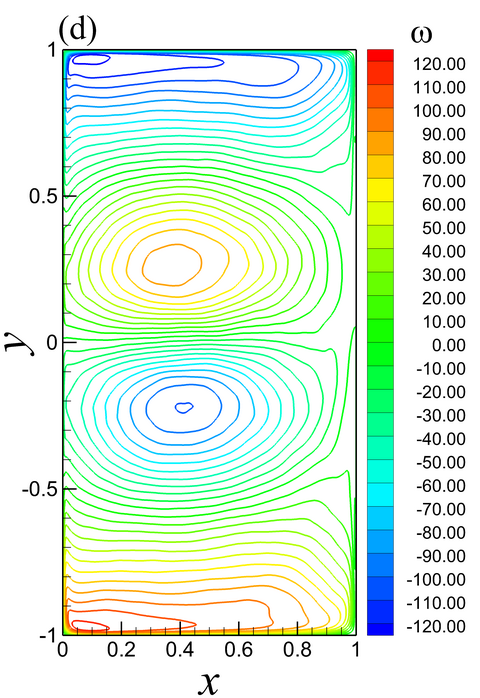}}
\subfigure{\includegraphics[width=\mywa]{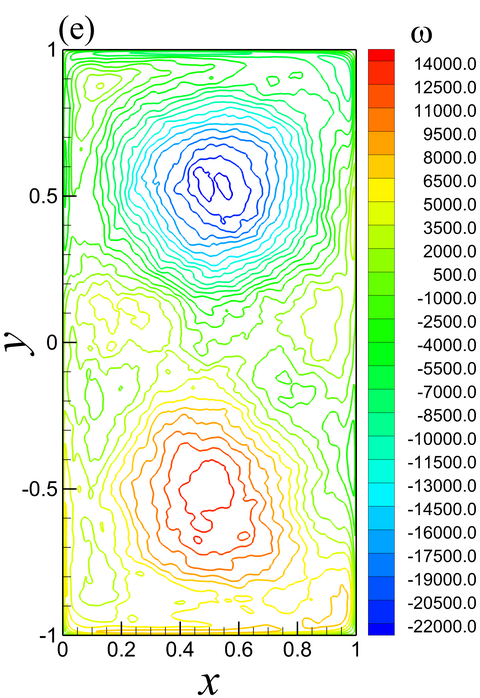}}
\subfigure{\includegraphics[width=\mywa]{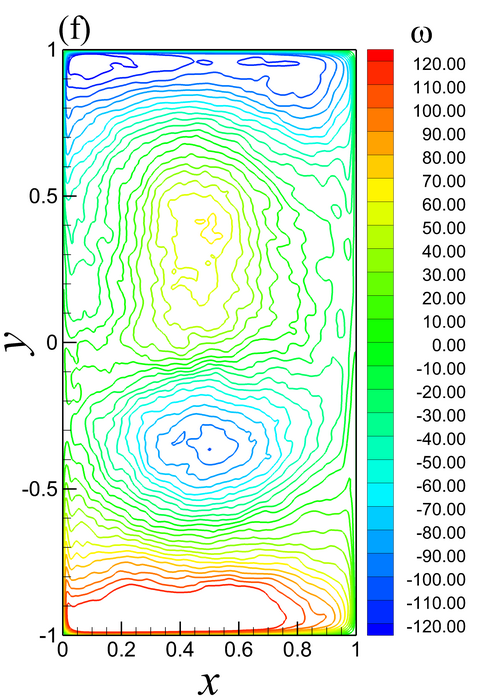}}
}
\caption{A comparison of the standard Galerkin approach (ROM) and the proposed ANN based stabilized approach (ROM-ANN) for $M=10$ modes. (a) $\psi$ by FOM; (b) $\psi$ by ROM; (c) $\psi$ by ROM-ANN; (d) $\omega$ by FOM; (e) $\omega$ by ROM; and (f) $\omega$ by ROM-ANN.}
\label{fig:u5}
\end{figure}

\begin{figure}[htbp]
\centering
\mbox{
\subfigure{\includegraphics[width=\mywa]{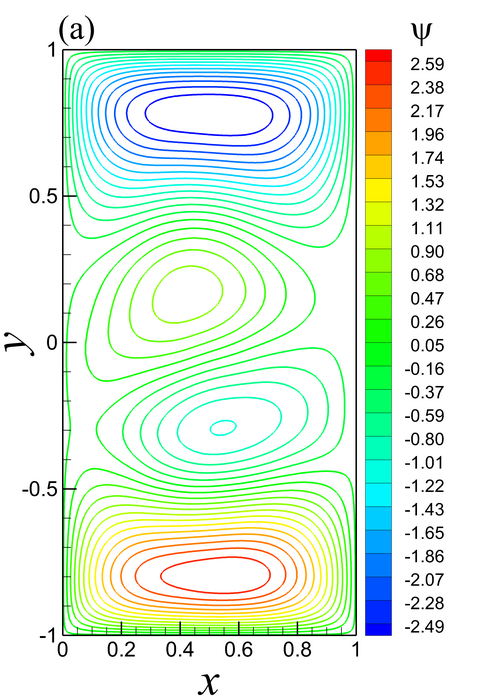}}
\subfigure{\includegraphics[width=\mywa]{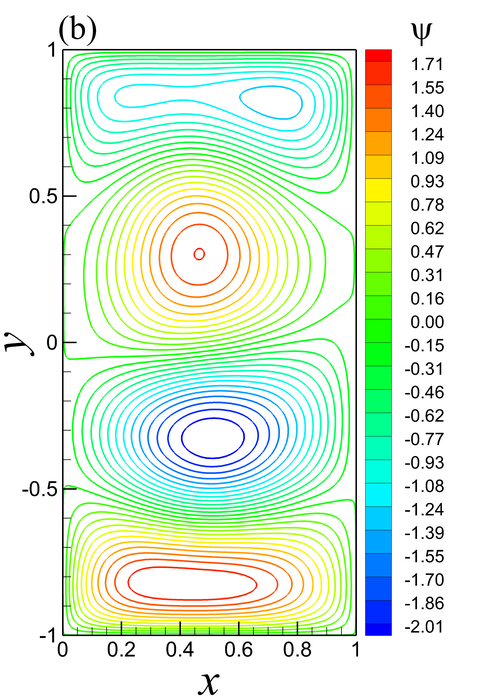}}
\subfigure{\includegraphics[width=\mywa]{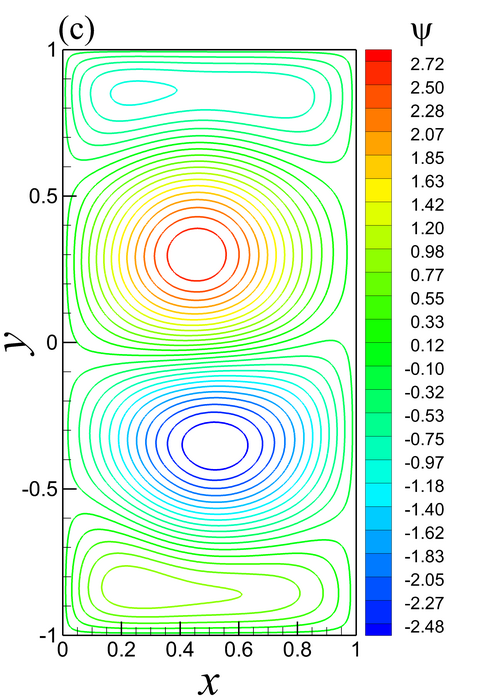}}
}\\
\mbox{
\subfigure{\includegraphics[width=\mywa]{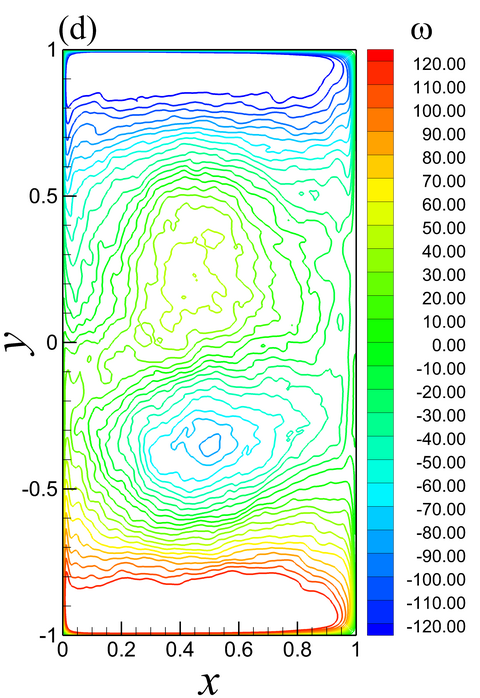}}
\subfigure{\includegraphics[width=\mywa]{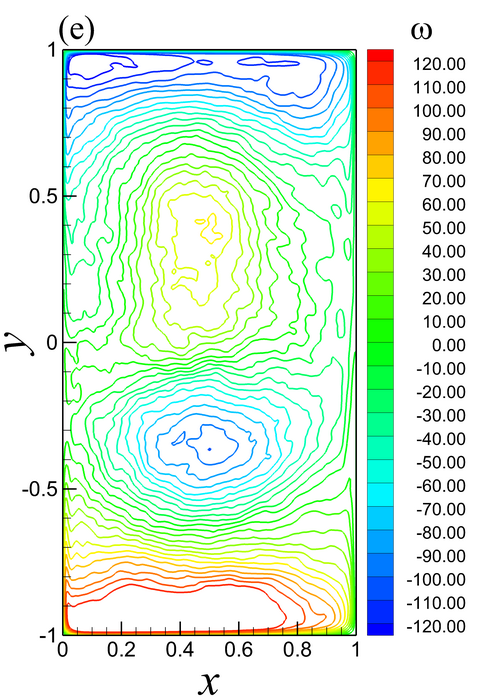}}
\subfigure{\includegraphics[width=\mywa]{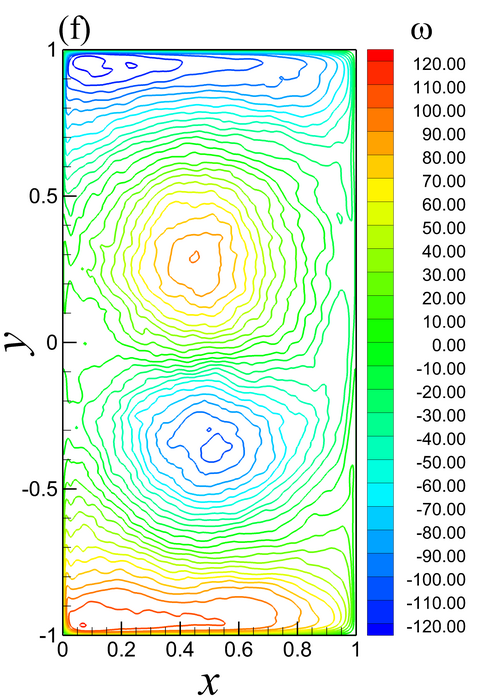}}
}
\caption{A sensitivity analysis with respect to the number of neurons in ELM using $M=10$ modes. (a) $\psi$ with $Q=20$ nodes; (b) $\psi$ with $Q=40$ nodes; (c) $\psi$ with $Q=60$ nodes; (d) $\omega$ with $Q=20$ nodes; (e) $\omega$ with $Q=40$ nodes; and (f) $\omega$ with $Q=60$ nodes.}
\label{fig:u6}
\end{figure}

Figure \ref{fig:u5} shows the performance of the proposed framework (i.e., ROM-ANN) against the standard Galerkin projection based ROM with $M=10$. Full order model (FOM) projections to reduced space are also shown for the purpose of comparison. It can easily be seen that the ELM stabilization reproduces the four-gyre pattern accurately as against the standard implementation of the ROM which fails to capture the pattern. This is observed for both streamfunction and vorticity contours. Figure \ref{fig:u6} shows a qualitative comparison of the effect of the number of neurons $Q$ where similar performance improvements are obtained for our choice of $Q = 20,40,60$ neurons. Table I shows a quantitative comparison of the improvement obtained by the proposed stabilization (for different neurons as well) against the standard ROM implementations with different number of modes. It is easily observed that the stabilization acts adequately in reproducing excellent agreement with full-order statistics at a very low number of retained modes. Note that these plots and tabulated statistics are all for the statistically steady state behavior of the QG problem in our assessment window (i.e., $t=10$ to $t=200$), which is beyond the training data window.

\begin{figure}[htbp]
\centering
\includegraphics[width=\mywd]{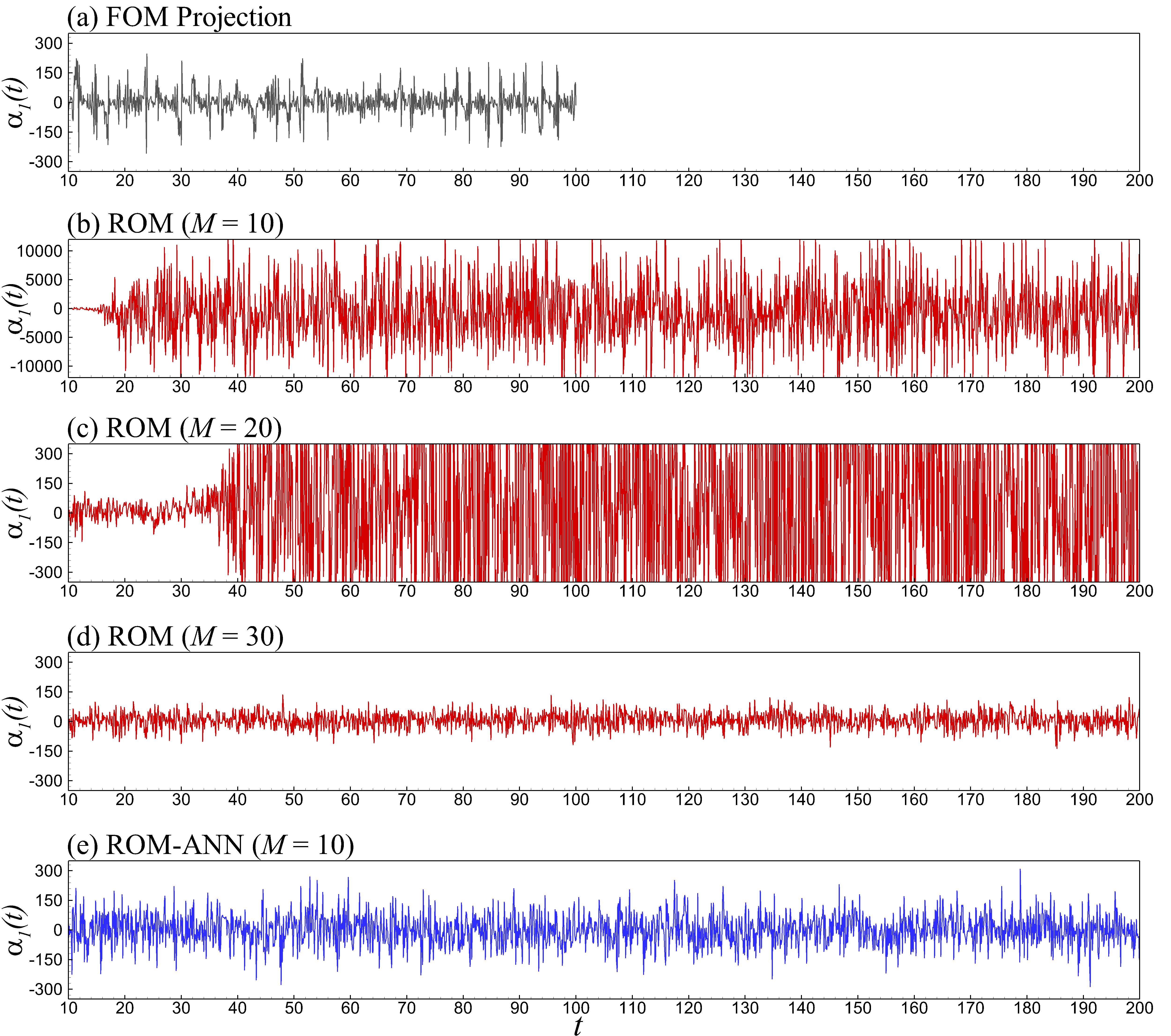}
\caption{Time series for the first temporal coefficient $\alpha_1(t)$.}
\label{fig:u7}
\end{figure}

Figure \ref{fig:u7} shows a comparison for the evolution of $\alpha_1$ through nondimensional time for both ROM and ROM-ANN implementations in comparison to the FOM projection. The ROM-ANN has a default $Q=40$ neurons in this example. It can clearly be seen that the use of the stabilization prevents the explosion of numerical instability in the coarse truncated ROMs with $M=10$ and $M=20$. At $M=30$ however, the first modal evolution shows a stable statistical steady state for the ROM. 


\begin{figure}[htbp]
\centering
\includegraphics[width=\mywd]{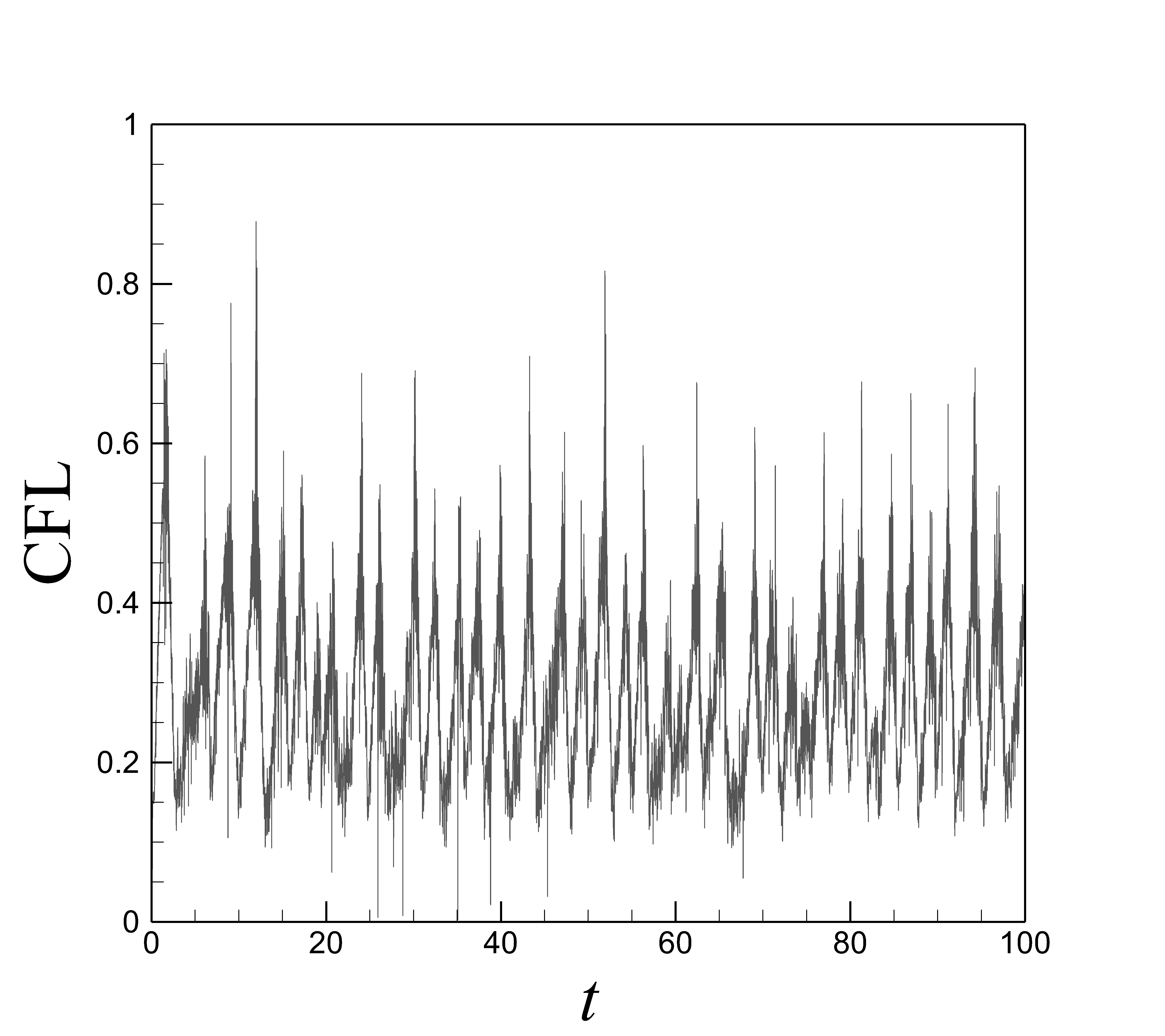}
\caption{Time series for the CFL criterion assuming a fixed time step of $\Delta t = 2.5 \times 10^{-5}$ for the FOM simulation.}
\label{fig:u9}
\end{figure}

\begin{figure}[htbp]
\centering
\mbox{
\subfigure{\includegraphics[width=\mywa]{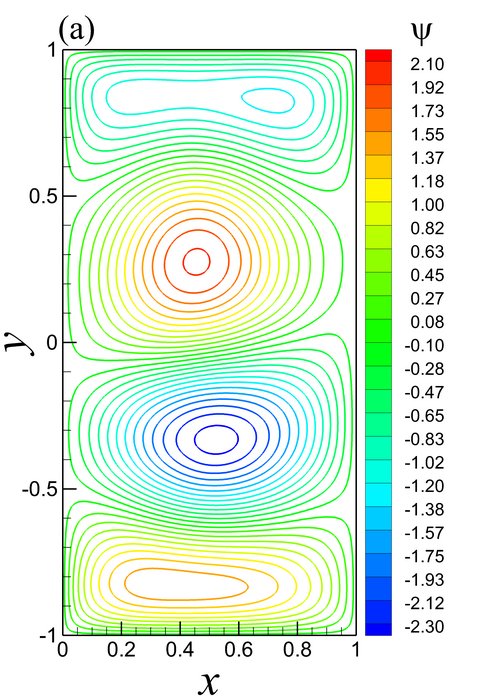}}
\subfigure{\includegraphics[width=\mywa]{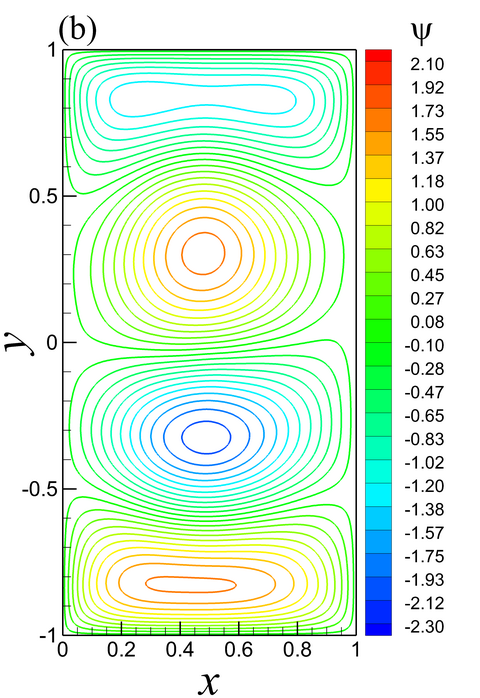}}
\subfigure{\includegraphics[width=\mywa]{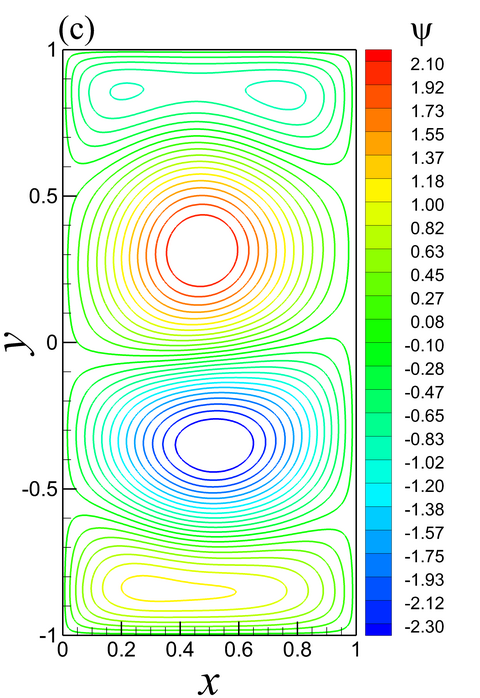}}
}\\
\mbox{
\subfigure{\includegraphics[width=\mywa]{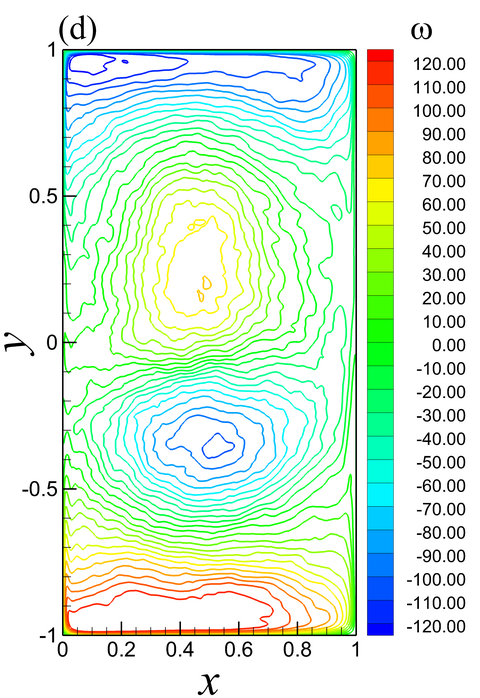}}
\subfigure{\includegraphics[width=\mywa]{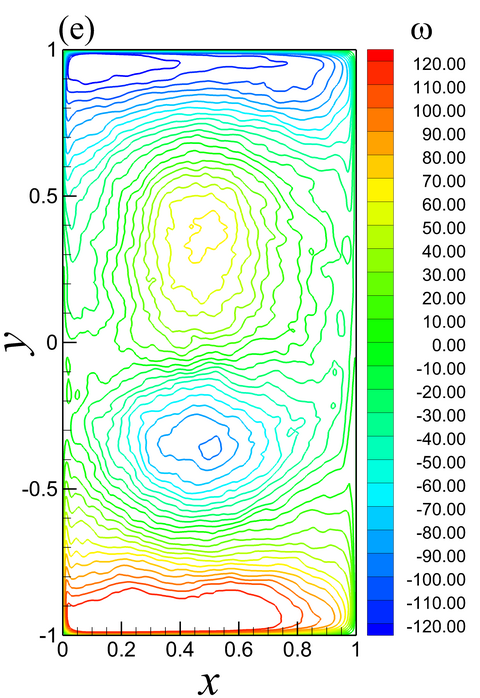}}
\subfigure{\includegraphics[width=\mywa]{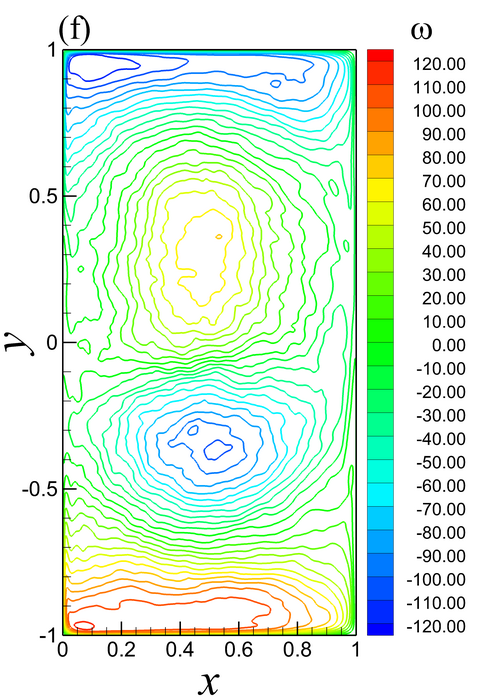}}
}
\caption{Mean streamfunction and vorticity contours obtained by the proposed ROM-ANN with various time steps. (a) $\psi$ with $\Delta t = 5 \times 10^{-4}$; (b) $\psi$ with $\Delta t = 2.5 \times 10^{-3}$; (c) $\psi$ with $\Delta t = 10^{-2}$; (d) $\omega$ with $\Delta t = 5 \times 10^{-4}$; (e) $\omega$ with $\Delta t = 2.5 \times 10^{-3}$; and (f) $\omega$ with $\Delta t = 10^{-2}$. Note that our ROM-ANN implementation uses $M=10$ and $Q = 40$.}
\label{fig:u10}
\end{figure}

\begin{figure}[htbp]
\centering
\includegraphics[width=\mywd]{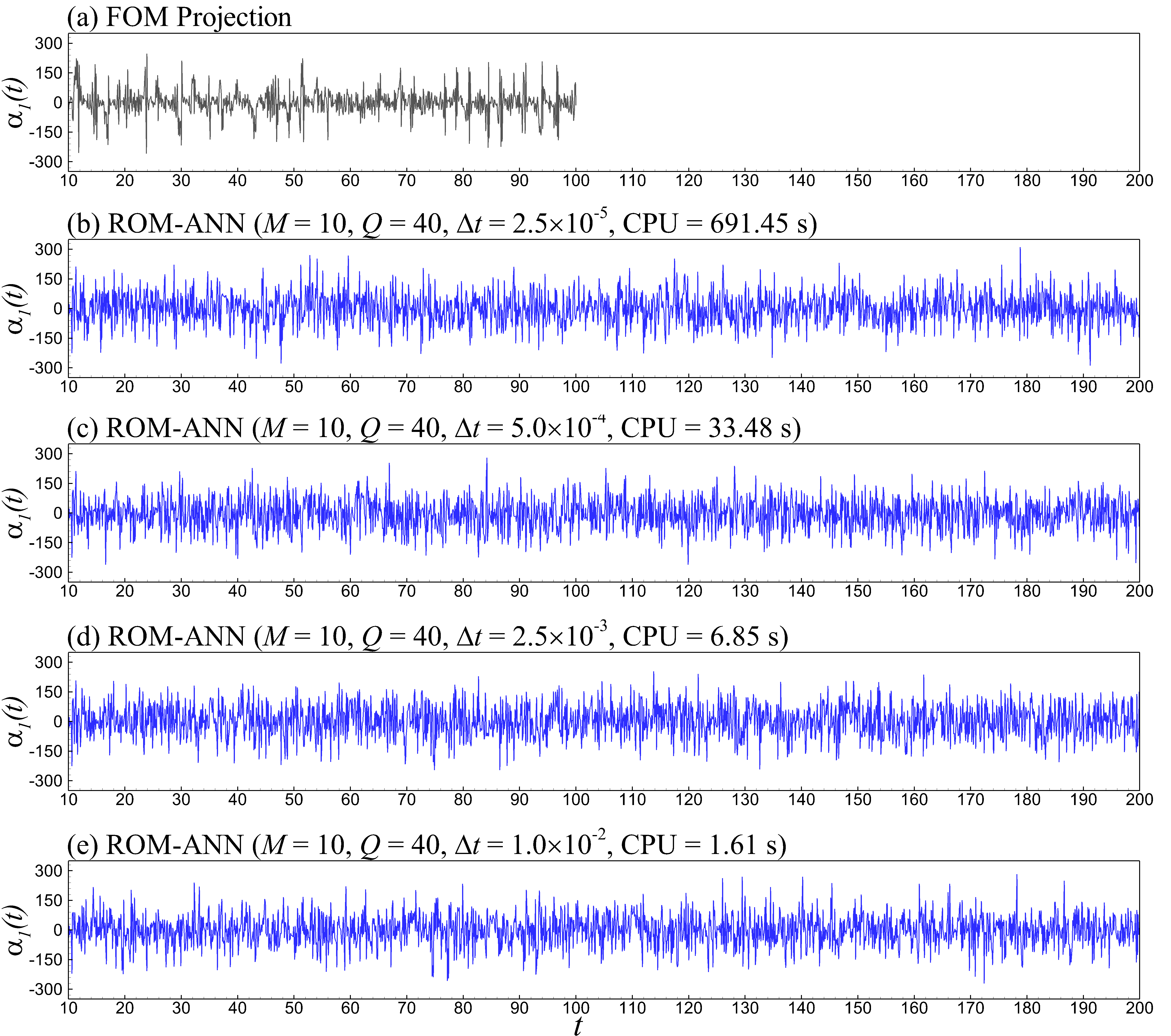}
\caption{Temporal mode evolution of $\alpha_1$ for various time steps.}
\label{fig:u11}
\end{figure}

Another benefit of the ROM-ANN mechanism over the standard ROM implementation is the possibility of using large time steps in the ordinary differential equation integrator. In the present study, a time step of $\Delta t=2.5 \times 10^{-5}$ was chosen for the FOM simulation to ensure a CFL criterion of less than 1.0 was always respected (as observed in the time series plot in Figure \ref{fig:u9}) due to the numerical stability of the numerical schemes. Figure \ref{fig:u10} shows the vorticity and streamfunction contours when our stabilized method (i.e., the ROM-ANN with $M=10$ and $Q=40$) is used with different time steps. It can be seen that a much larger time step of $\Delta t = 10^{-2}$ can be effectively used to obtain statistically accurate results without any divergence. Thus our proposed ANN based eddy viscosity stabilization is ideally suited to a fast prediction of the underlying dynamics. Figure \ref{fig:u11} shows the evolution of the first temporal coefficient $\alpha_1(t)$ for the aforementioned ROM-ANN framework where it is seen that very high values of the time step do not affect the statistical viability of the stabilized ROM and leads to an excellent reduction in computational expense (the largest time step provides excellent results at a CPU time of 1.61 seconds in comparison to approximately 700 seconds for the default time step which is required for the FOM). We also note that the FOM required 195.4 hours CPU time to complete the forward simulation between $t=10$ and $t=100$. 

\begin{figure}[htbp]
\centering
\mbox{
\subfigure{\includegraphics[width=\mywa]{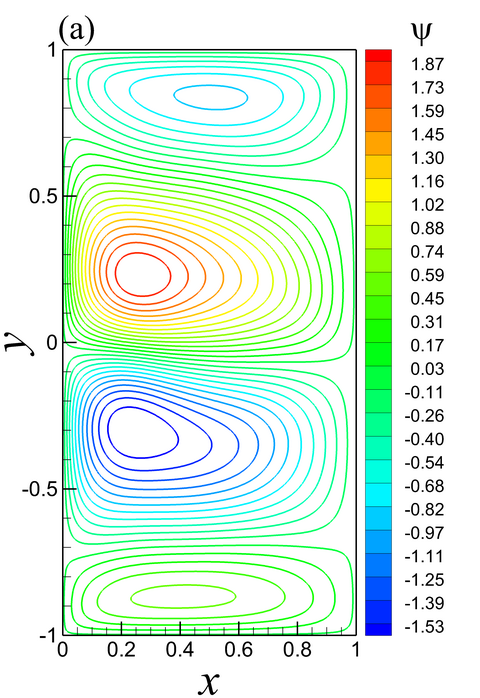}}
\subfigure{\includegraphics[width=\mywa]{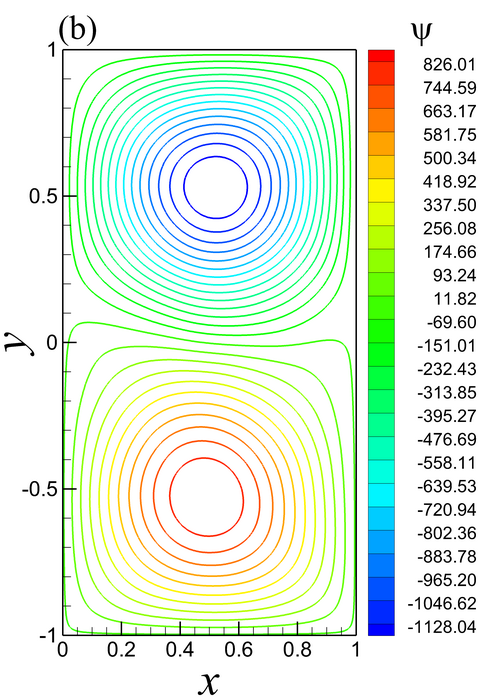}}
\subfigure{\includegraphics[width=\mywa]{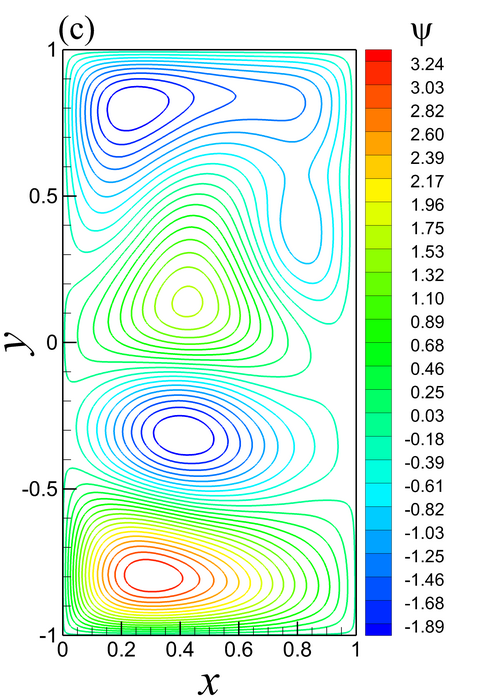}}
}\\
\mbox{
\subfigure{\includegraphics[width=\mywa]{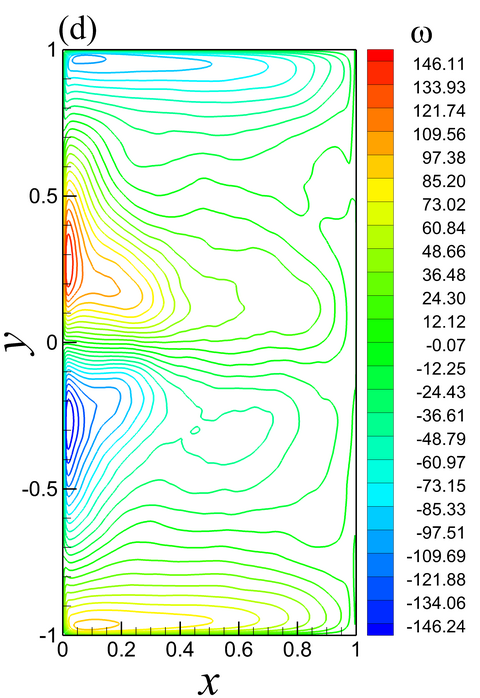}}
\subfigure{\includegraphics[width=\mywa]{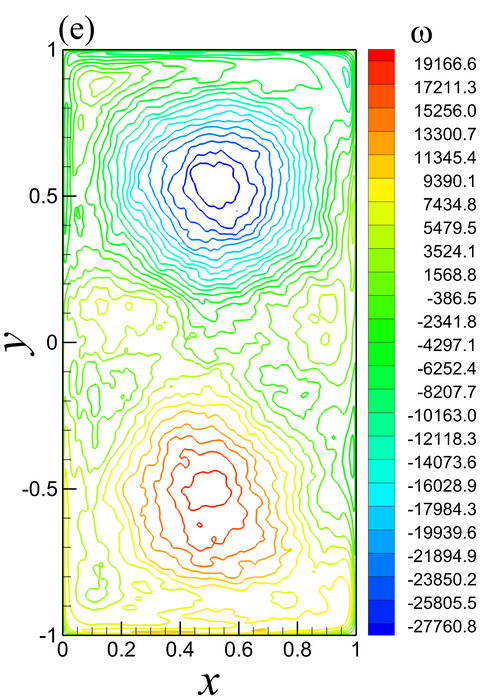}}
\subfigure{\includegraphics[width=\mywa]{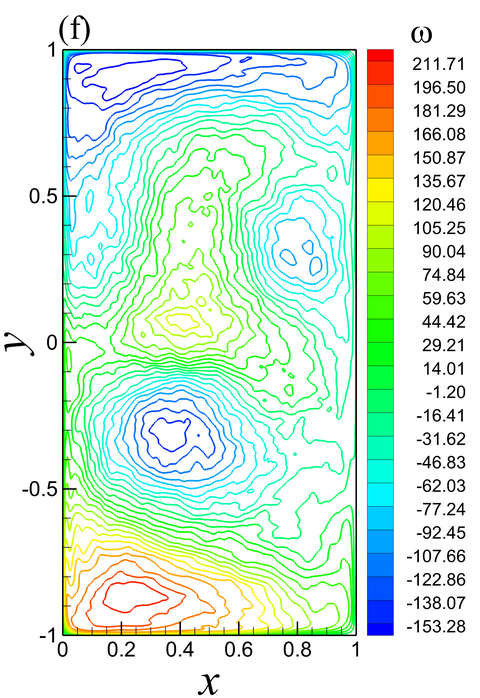}}
}
\caption{A comparison of the standard Galerkin approach (ROM) and the proposed ANN based stabilized approach (ROM-ANN) for $M=10$ modes at $Re=200$ and $Ro=1.6 \times 10^{-3}$ where the training data has been extracted from the snapshots at $Re=450$ and $Ro=3.6 \times 10^{-3}$. (a) $\psi$ by FOM; (b) $\psi$ by ROM; (c) $\psi$ by ROM-ANN; (d) $\omega$ by FOM; (e) $\omega$ by ROM; and (f) $\omega$ by ROM-ANN. Note that the ROM-ANN uses $Q=40$ hidden nodes.}
\label{fig:aa1}
\end{figure}
\begin{figure}[htbp]
\centering
\includegraphics[width=\mywd]{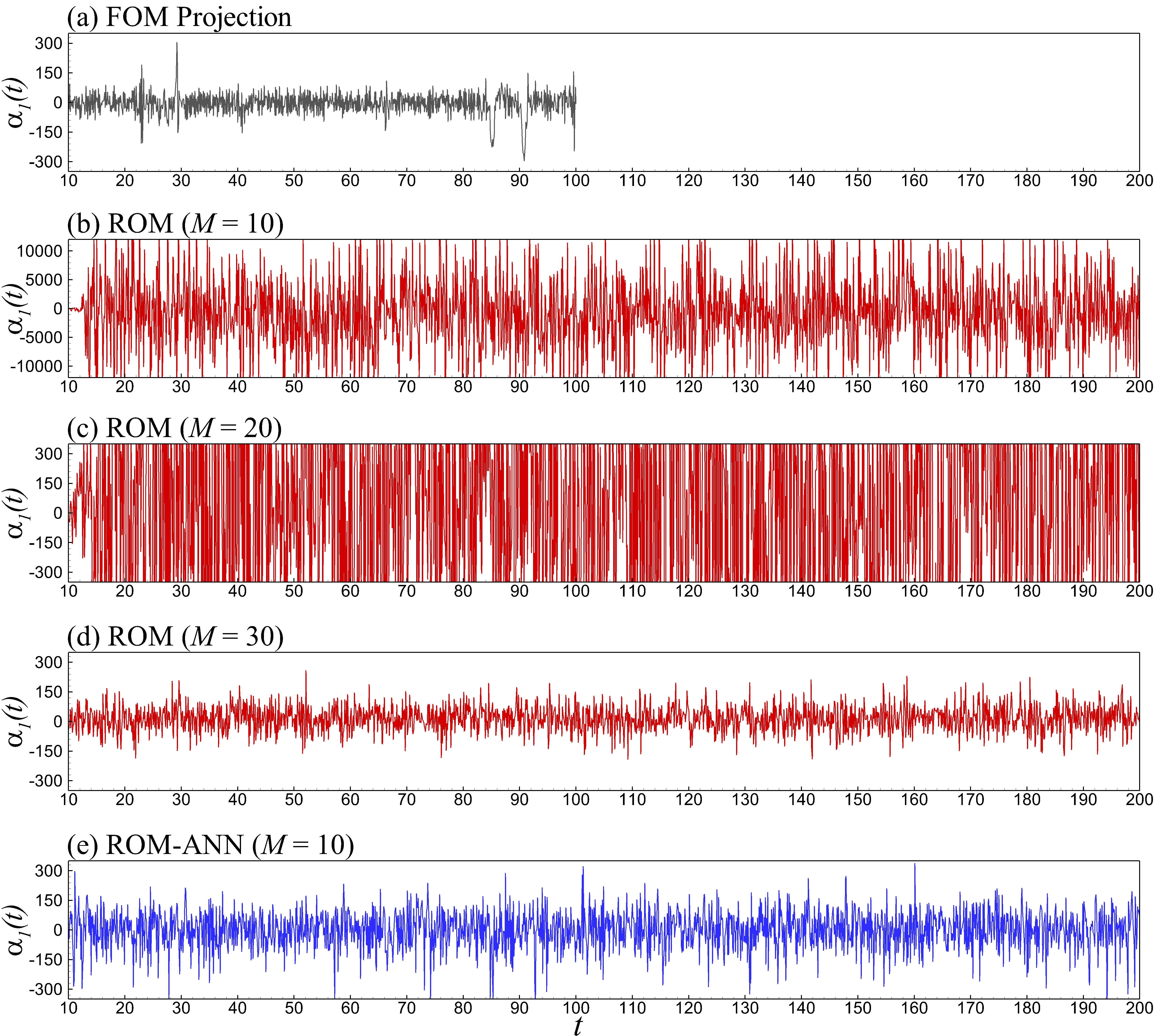}
\caption{Time series for of first temporal coefficient $\alpha_1(t)$ for the out-of-sample forecast. Predictive performance is shown for $Re=200$ and $Ro=1.6 \times 10^{-3}$ while the training has been performed using the data generated at $Re=450$ and $Ro=3.6 \times 10^{-3}$.}
\label{fig:aa2}
\end{figure}

Finally, we perform an out-of-sample a-posteriori analysis considering different physical parameters than those used in the training data. As explained earlier the physical model parameters are $Re=450$ and $Ro=3.6 \times 10^{-3}$ to generate our data snapshots (therefore POD basis functions) and the supervised training data set for ELM. Using the same ELM network and POD basis functions, Figure \ref{fig:aa1} compares the predictive performance of the models at $Re=200$ and $Ro=1.6 \times 10^{-3}$, a distinct test set-up from the training data. A new FOM simulation is performed for our assessments (i.e., required about 200.6 hours CPU time). It is clear that the ROM-ANN captures the main dynamics requiring only the order of seconds CPU time for the simulation. Time series of $\alpha_1$ are also illustrated in Figure \ref{fig:aa2}. Similar to the in-sample case, the standard Galerkin ROM approach cannot capture the correct dynamics for $M=10$ or $M=20$ modes. On the other hand, the proposed stabilized ROM-ANN approach captures the underlying four-gyre dynamics and provides significantly accurate results for using $M=10$ modes. Our assessments conclude that the proposed architecture is robust in providing a reliable mode dependent damping coefficient for the out-of-sample forecasting.

\section{Summary and Conclusions}
\label{sec:con}

In this paper, we have studied the feasibility of using a machine learning framework to stabilize projection based ROMs for solving a forced-dissipative general circulation problem. We construct an SLFN to predict modal eddy viscosity coefficients dynamically. Our approach can be considered semi non-intrusive (without the need for an online access to the FOM for the ROM prediction), since the ANN architecture only requires reduced order space quantities to predict the stabilization term. A regularized ELM approach is used for training where we use the same data snapshots as we used for generating the POD basis functions. In that sense, there are two data-driven components to this research: high fidelity snapshots of data from DNS are utilized not just for POD basis synthesis but also for training our machine learning framework utilized for a-posteriori stabilization of the ROM-ANN. Both in-sample and out-of-sample simulation results indicate that the utilization of the proposed framework lets the user deploy an extremely truncated system without losing any statistical fidelity. Also, time steps much larger than those necessary for FOM forward simulations may be utilized in the ROM-ANN thus leading to exceptional computational performance. We conclude that the method presented in this paper is robust enough to stabilize ROMs dynamically and satisfies the dual demands of statistical accuracy as well as low computational expense in surrogate forward predictions in long-term evolution of geophysical turbulent flows.

\section*{Acknowledgment}
The computing for this project was performed by using resources from the High Performance Computing Center (HPCC) at Oklahoma State University.

\newpage

\bibliography{reference}

\begin{thebibliography}{49}
\expandafter\ifx\csname natexlab\endcsname\relax\def\natexlab#1{#1}\fi
\expandafter\ifx\csname bibnamefont\endcsname\relax
  \def\bibnamefont#1{#1}\fi
\expandafter\ifx\csname bibfnamefont\endcsname\relax
  \def\bibfnamefont#1{#1}\fi
\expandafter\ifx\csname citenamefont\endcsname\relax
  \def\citenamefont#1{#1}\fi
\expandafter\ifx\csname url\endcsname\relax
  \def\url#1{\texttt{#1}}\fi
\expandafter\ifx\csname urlprefix\endcsname\relax\def\urlprefix{URL }\fi
\providecommand{\bibinfo}[2]{#2}
\providecommand{\eprint}[2][]{\url{#2}}

\bibitem[{\citenamefont{Brunton and Noack}(2015)}]{brunton2015closed}
\bibinfo{author}{\bibfnamefont{S.~L.} \bibnamefont{Brunton}} \bibnamefont{and}
  \bibinfo{author}{\bibfnamefont{B.~R.} \bibnamefont{Noack}},
  {``}\bibinfo{title}{Closed-loop turbulence control: Progress and
  challenges},{''} \bibinfo{journal}{Applied Mechanics Reviews}
  \textbf{\bibinfo{volume}{67}}, \bibinfo{pages}{050801}
  (\bibinfo{year}{2015}).

\bibitem[{\citenamefont{Daescu and Navon}(2007)}]{daescu2007efficiency}
\bibinfo{author}{\bibfnamefont{D.~N.} \bibnamefont{Daescu}} \bibnamefont{and}
  \bibinfo{author}{\bibfnamefont{I.~M.} \bibnamefont{Navon}},
  {``}\bibinfo{title}{{Efficiency of a POD-based reduced second-order adjoint
  model in 4D-Var data assimilation}},{''} \bibinfo{journal}{International
  Journal for Numerical Methods in Fluids} \textbf{\bibinfo{volume}{53}},
  \bibinfo{pages}{985} (\bibinfo{year}{2007}).

\bibitem[{\citenamefont{Cao et~al.}(2007)\citenamefont{Cao, Zhu, Navon, and
  Luo}}]{cao2007reduced}
\bibinfo{author}{\bibfnamefont{Y.}~\bibnamefont{Cao}},
  \bibinfo{author}{\bibfnamefont{J.}~\bibnamefont{Zhu}},
  \bibinfo{author}{\bibfnamefont{I.~M.} \bibnamefont{Navon}}, \bibnamefont{and}
  \bibinfo{author}{\bibfnamefont{Z.}~\bibnamefont{Luo}}, {``}\bibinfo{title}{A
  reduced-order approach to four-dimensional variational data assimilation
  using proper orthogonal decomposition},{''} \bibinfo{journal}{International
  Journal for Numerical Methods in Fluids} \textbf{\bibinfo{volume}{53}},
  \bibinfo{pages}{1571} (\bibinfo{year}{2007}).

\bibitem[{\citenamefont{Daescu and Navon}(2008)}]{daescu2008dual}
\bibinfo{author}{\bibfnamefont{D.}~\bibnamefont{Daescu}} \bibnamefont{and}
  \bibinfo{author}{\bibfnamefont{I.}~\bibnamefont{Navon}},
  {``}\bibinfo{title}{{A dual-weighted approach to order reduction in 4DVAR
  data assimilation}},{''} \bibinfo{journal}{Monthly Weather Review}
  \textbf{\bibinfo{volume}{136}}, \bibinfo{pages}{1026} (\bibinfo{year}{2008}).

\bibitem[{\citenamefont{Benner et~al.}(2015)\citenamefont{Benner, Gugercin, and
  Willcox}}]{benner2015survey}
\bibinfo{author}{\bibfnamefont{P.}~\bibnamefont{Benner}},
  \bibinfo{author}{\bibfnamefont{S.}~\bibnamefont{Gugercin}}, \bibnamefont{and}
  \bibinfo{author}{\bibfnamefont{K.}~\bibnamefont{Willcox}},
  {``}\bibinfo{title}{A survey of projection-based model reduction methods for
  parametric dynamical systems},{''} \bibinfo{journal}{SIAM Review}
  \textbf{\bibinfo{volume}{57}}, \bibinfo{pages}{483} (\bibinfo{year}{2015}).

\bibitem[{\citenamefont{Rowley and Dawson}(2017)}]{rowley2017model}
\bibinfo{author}{\bibfnamefont{C.~W.} \bibnamefont{Rowley}} \bibnamefont{and}
  \bibinfo{author}{\bibfnamefont{S.~T.} \bibnamefont{Dawson}},
  {``}\bibinfo{title}{Model reduction for flow analysis and control},{''}
  \bibinfo{journal}{Annual Review of Fluid Mechanics}
  \textbf{\bibinfo{volume}{49}}, \bibinfo{pages}{387} (\bibinfo{year}{2017}).

\bibitem[{\citenamefont{Taira et~al.}(2017)\citenamefont{Taira, Brunton,
  Dawson, Rowley, Colonius, McKeon, Schmidt, Gordeyev, Theofilis, and
  Ukeiley}}]{taira2017modal}
\bibinfo{author}{\bibfnamefont{K.}~\bibnamefont{Taira}},
  \bibinfo{author}{\bibfnamefont{S.~L.} \bibnamefont{Brunton}},
  \bibinfo{author}{\bibfnamefont{S.}~\bibnamefont{Dawson}},
  \bibinfo{author}{\bibfnamefont{C.~W.} \bibnamefont{Rowley}},
  \bibinfo{author}{\bibfnamefont{T.}~\bibnamefont{Colonius}},
  \bibinfo{author}{\bibfnamefont{B.~J.} \bibnamefont{McKeon}},
  \bibinfo{author}{\bibfnamefont{O.~T.} \bibnamefont{Schmidt}},
  \bibinfo{author}{\bibfnamefont{S.}~\bibnamefont{Gordeyev}},
  \bibinfo{author}{\bibfnamefont{V.}~\bibnamefont{Theofilis}},
  \bibnamefont{and} \bibinfo{author}{\bibfnamefont{L.~S.}
  \bibnamefont{Ukeiley}}, {``}\bibinfo{title}{Modal analysis of fluid flows: An
  overview},{''} \bibinfo{journal}{AIAA Journal} \textbf{\bibinfo{volume}{55}},
  \bibinfo{pages}{4013} (\bibinfo{year}{2017}).

\bibitem[{\citenamefont{Holmes et~al.}(1998)\citenamefont{Holmes, Lumley, and
  Berkooz}}]{holmes1998turbulence}
\bibinfo{author}{\bibfnamefont{P.}~\bibnamefont{Holmes}},
  \bibinfo{author}{\bibfnamefont{J.~L.} \bibnamefont{Lumley}},
  \bibnamefont{and} \bibinfo{author}{\bibfnamefont{G.}~\bibnamefont{Berkooz}},
  \emph{\bibinfo{title}{Turbulence, coherent structures, dynamical systems and
  symmetry}} (\bibinfo{publisher}{Cambridge University Press},
  \bibinfo{year}{1998}).

\bibitem[{\citenamefont{Ito and Ravindran}(1998)}]{ito1998reduced}
\bibinfo{author}{\bibfnamefont{K.}~\bibnamefont{Ito}} \bibnamefont{and}
  \bibinfo{author}{\bibfnamefont{S.}~\bibnamefont{Ravindran}},
  {``}\bibinfo{title}{A reduced-order method for simulation and control of
  fluid flows},{''} \bibinfo{journal}{Journal of Computational Physics}
  \textbf{\bibinfo{volume}{143}}, \bibinfo{pages}{403} (\bibinfo{year}{1998}).

\bibitem[{\citenamefont{Iollo et~al.}(2000)\citenamefont{Iollo, Lanteri, and
  D{\'e}sid{\'e}ri}}]{iollo2000stability}
\bibinfo{author}{\bibfnamefont{A.}~\bibnamefont{Iollo}},
  \bibinfo{author}{\bibfnamefont{S.}~\bibnamefont{Lanteri}}, \bibnamefont{and}
  \bibinfo{author}{\bibfnamefont{J.-A.} \bibnamefont{D{\'e}sid{\'e}ri}},
  {``}\bibinfo{title}{Stability properties of
  \uppercase{POD}--\uppercase{G}alerkin approximations for the compressible
  \uppercase{N}avier--\uppercase{S}tokes equations},{''}
  \bibinfo{journal}{Theoretical and Computational Fluid Dynamics}
  \textbf{\bibinfo{volume}{13}}, \bibinfo{pages}{377} (\bibinfo{year}{2000}).

\bibitem[{\citenamefont{Noack et~al.}(2003)\citenamefont{Noack, Afanasiev,
  Morzynski, Tadmor, and Thiele}}]{noack2003hierarchy}
\bibinfo{author}{\bibfnamefont{B.~R.} \bibnamefont{Noack}},
  \bibinfo{author}{\bibfnamefont{K.}~\bibnamefont{Afanasiev}},
  \bibinfo{author}{\bibfnamefont{M.}~\bibnamefont{Morzynski}},
  \bibinfo{author}{\bibfnamefont{G.}~\bibnamefont{Tadmor}}, \bibnamefont{and}
  \bibinfo{author}{\bibfnamefont{F.}~\bibnamefont{Thiele}},
  {``}\bibinfo{title}{A hierarchy of low-dimensional models for the transient
  and post-transient cylinder wake},{''} \bibinfo{journal}{Journal of Fluid
  Mechanics} \textbf{\bibinfo{volume}{497}}, \bibinfo{pages}{335}
  (\bibinfo{year}{2003}).

\bibitem[{\citenamefont{Narasimha}(2011)}]{narasimha2011kosambi}
\bibinfo{author}{\bibfnamefont{R.}~\bibnamefont{Narasimha}},
  {``}\bibinfo{title}{Kosambi and proper orthogonal decomposition},{''}
  \bibinfo{journal}{Resonance} \textbf{\bibinfo{volume}{16}},
  \bibinfo{pages}{574} (\bibinfo{year}{2011}).

\bibitem[{\citenamefont{Couplet et~al.}(2003)\citenamefont{Couplet, Sagaut, and
  Basdevant}}]{couplet2003intermodal}
\bibinfo{author}{\bibfnamefont{M.}~\bibnamefont{Couplet}},
  \bibinfo{author}{\bibfnamefont{P.}~\bibnamefont{Sagaut}}, \bibnamefont{and}
  \bibinfo{author}{\bibfnamefont{C.}~\bibnamefont{Basdevant}},
  {``}\bibinfo{title}{Intermodal energy transfers in a proper orthogonal
  decomposition-\uppercase{G}alerkin representation of a turbulent separated
  flow},{''} \bibinfo{journal}{Journal of Fluid Mechanics}
  \textbf{\bibinfo{volume}{491}}, \bibinfo{pages}{275} (\bibinfo{year}{2003}).

\bibitem[{\citenamefont{Kalb and Deane}(2007)}]{kalb2007intrinsic}
\bibinfo{author}{\bibfnamefont{V.~L.} \bibnamefont{Kalb}} \bibnamefont{and}
  \bibinfo{author}{\bibfnamefont{A.~E.} \bibnamefont{Deane}},
  {``}\bibinfo{title}{An intrinsic stabilization scheme for proper orthogonal
  decomposition based low-dimensional models},{''} \bibinfo{journal}{Physics of
  fluids} \textbf{\bibinfo{volume}{19}}, \bibinfo{pages}{054106}
  (\bibinfo{year}{2007}).

\bibitem[{\citenamefont{Bergmann et~al.}(2009)\citenamefont{Bergmann, Bruneau,
  and Iollo}}]{bergmann2009enablers}
\bibinfo{author}{\bibfnamefont{M.}~\bibnamefont{Bergmann}},
  \bibinfo{author}{\bibfnamefont{C.-H.} \bibnamefont{Bruneau}},
  \bibnamefont{and} \bibinfo{author}{\bibfnamefont{A.}~\bibnamefont{Iollo}},
  {``}\bibinfo{title}{Enablers for robust \uppercase{POD} models},{''}
  \bibinfo{journal}{Journal of Computational Physics}
  \textbf{\bibinfo{volume}{228}}, \bibinfo{pages}{516} (\bibinfo{year}{2009}).

\bibitem[{\citenamefont{Wang et~al.}(2012)\citenamefont{Wang, Akhtar,
  Borggaard, and Iliescu}}]{wang2012proper}
\bibinfo{author}{\bibfnamefont{Z.}~\bibnamefont{Wang}},
  \bibinfo{author}{\bibfnamefont{I.}~\bibnamefont{Akhtar}},
  \bibinfo{author}{\bibfnamefont{J.}~\bibnamefont{Borggaard}},
  \bibnamefont{and} \bibinfo{author}{\bibfnamefont{T.}~\bibnamefont{Iliescu}},
  {``}\bibinfo{title}{Proper orthogonal decomposition closure models for
  turbulent flows: a numerical comparison},{''} \bibinfo{journal}{Computer
  Methods in Applied Mechanics and Engineering}
  \textbf{\bibinfo{volume}{237--240}}, \bibinfo{pages}{10}
  (\bibinfo{year}{2012}).

\bibitem[{\citenamefont{Baiges et~al.}(2015)\citenamefont{Baiges, Codina, and
  Idelsohn}}]{baiges2015reduced}
\bibinfo{author}{\bibfnamefont{J.}~\bibnamefont{Baiges}},
  \bibinfo{author}{\bibfnamefont{R.}~\bibnamefont{Codina}}, \bibnamefont{and}
  \bibinfo{author}{\bibfnamefont{S.}~\bibnamefont{Idelsohn}},
  {``}\bibinfo{title}{Reduced-order subscales for POD models},{''}
  \bibinfo{journal}{Computer Methods in Applied Mechanics and Engineering}
  \textbf{\bibinfo{volume}{291}}, \bibinfo{pages}{173} (\bibinfo{year}{2015}).

\bibitem[{\citenamefont{Xie et~al.}(2017)\citenamefont{Xie, Wells, Wang, and
  Iliescu}}]{xie2017approximate}
\bibinfo{author}{\bibfnamefont{X.}~\bibnamefont{Xie}},
  \bibinfo{author}{\bibfnamefont{D.}~\bibnamefont{Wells}},
  \bibinfo{author}{\bibfnamefont{Z.}~\bibnamefont{Wang}}, \bibnamefont{and}
  \bibinfo{author}{\bibfnamefont{T.}~\bibnamefont{Iliescu}},
  {``}\bibinfo{title}{Approximate deconvolution reduced order modeling},{''}
  \bibinfo{journal}{Computer Methods in Applied Mechanics and Engineering}
  \textbf{\bibinfo{volume}{313}}, \bibinfo{pages}{512} (\bibinfo{year}{2017}).

\bibitem[{\citenamefont{Akhtar et~al.}(2012)\citenamefont{Akhtar, Wang,
  Borggaard, and Iliescu}}]{akhtar2012new}
\bibinfo{author}{\bibfnamefont{I.}~\bibnamefont{Akhtar}},
  \bibinfo{author}{\bibfnamefont{Z.}~\bibnamefont{Wang}},
  \bibinfo{author}{\bibfnamefont{J.}~\bibnamefont{Borggaard}},
  \bibnamefont{and} \bibinfo{author}{\bibfnamefont{T.}~\bibnamefont{Iliescu}},
  {``}\bibinfo{title}{A new closure strategy for proper orthogonal
  decomposition reduced-order models},{''} \bibinfo{journal}{Journal of
  Computational and Nonlinear Dynamics} \textbf{\bibinfo{volume}{7}},
  \bibinfo{pages}{034503} (\bibinfo{year}{2012}).

\bibitem[{\citenamefont{San and Iliescu}(2014)}]{san2014proper}
\bibinfo{author}{\bibfnamefont{O.}~\bibnamefont{San}} \bibnamefont{and}
  \bibinfo{author}{\bibfnamefont{T.}~\bibnamefont{Iliescu}},
  {``}\bibinfo{title}{{Proper orthogonal decomposition closure models for fluid
  flows: Burgers equation}},{''} \bibinfo{journal}{International Journal of
  Numerical Analysis and Modeling, Series B} \textbf{\bibinfo{volume}{5}},
  \bibinfo{pages}{217} (\bibinfo{year}{2014}).

\bibitem[{\citenamefont{San and Iliescu}(2015)}]{san2015stabilized}
\bibinfo{author}{\bibfnamefont{O.}~\bibnamefont{San}} \bibnamefont{and}
  \bibinfo{author}{\bibfnamefont{T.}~\bibnamefont{Iliescu}},
  {``}\bibinfo{title}{A stabilized proper orthogonal decomposition
  reduced-order model for large scale quasigeostrophic ocean circulation},{''}
  \bibinfo{journal}{Advances in Computational Mathematics}
  \textbf{\bibinfo{volume}{41}}, \bibinfo{pages}{1289} (\bibinfo{year}{2015}).

\bibitem[{\citenamefont{Narayanan et~al.}(1999)\citenamefont{Narayanan,
  Khibnik, Jacobson, Kevrekedis, Rico-Martinez, and Lust}}]{narayanan1999low}
\bibinfo{author}{\bibfnamefont{S.}~\bibnamefont{Narayanan}},
  \bibinfo{author}{\bibfnamefont{A.}~\bibnamefont{Khibnik}},
  \bibinfo{author}{\bibfnamefont{C.}~\bibnamefont{Jacobson}},
  \bibinfo{author}{\bibfnamefont{Y.}~\bibnamefont{Kevrekedis}},
  \bibinfo{author}{\bibfnamefont{R.}~\bibnamefont{Rico-Martinez}},
  \bibnamefont{and} \bibinfo{author}{\bibfnamefont{K.}~\bibnamefont{Lust}}, in
  \emph{\bibinfo{booktitle}{Control Applications, 1999. Proceedings of the 1999
  IEEE International Conference on}} (\bibinfo{organization}{IEEE},
  \bibinfo{year}{1999}), vol.~\bibinfo{volume}{2}, pp.
  \bibinfo{pages}{1151--1156}.

\bibitem[{\citenamefont{Sahan et~al.}(1997)\citenamefont{Sahan, Koc-Sahan,
  Albin, and Liakopoulos}}]{sahan1997artificial}
\bibinfo{author}{\bibfnamefont{R.}~\bibnamefont{Sahan}},
  \bibinfo{author}{\bibfnamefont{N.}~\bibnamefont{Koc-Sahan}},
  \bibinfo{author}{\bibfnamefont{D.}~\bibnamefont{Albin}}, \bibnamefont{and}
  \bibinfo{author}{\bibfnamefont{A.}~\bibnamefont{Liakopoulos}}, in
  \emph{\bibinfo{booktitle}{Proceedings of the 1997 IEEE International
  Conference on Control Applications,}} (\bibinfo{organization}{IEEE},
  \bibinfo{year}{1997}), pp. \bibinfo{pages}{359--364}.

\bibitem[{\citenamefont{Moosavi et~al.}(2015)\citenamefont{Moosavi, Stefanescu,
  and Sandu}}]{moosavi2015efficient}
\bibinfo{author}{\bibfnamefont{A.}~\bibnamefont{Moosavi}},
  \bibinfo{author}{\bibfnamefont{R.}~\bibnamefont{Stefanescu}},
  \bibnamefont{and} \bibinfo{author}{\bibfnamefont{A.}~\bibnamefont{Sandu}},
  {``}\bibinfo{title}{Efficient {C}onstruction of {L}ocal {P}arametric
  {R}educed {O}rder {M}odels {U}sing {M}achine {L}earning {T}echniques},{''}
  \bibinfo{journal}{arXiv preprint arXiv:1511.02909}  (\bibinfo{year}{2015}).

\bibitem[{\citenamefont{San and Maulik}(2018)}]{san2018neural}
\bibinfo{author}{\bibfnamefont{O.}~\bibnamefont{San}} \bibnamefont{and}
  \bibinfo{author}{\bibfnamefont{R.}~\bibnamefont{Maulik}},
  {``}\bibinfo{title}{Neural network closures for nonlinear model order
  reduction},{''} \bibinfo{journal}{Advances in Computational Mathematics,
  DOI:10.1007/s10444-018-9590-z}  (\bibinfo{year}{2018}).

\bibitem[{\citenamefont{Gillies}(1998)}]{gillies1998low}
\bibinfo{author}{\bibfnamefont{E.}~\bibnamefont{Gillies}},
  {``}\bibinfo{title}{Low-dimensional control of the circular cylinder
  wake},{''} \bibinfo{journal}{Journal of Fluid Mechanics}
  \textbf{\bibinfo{volume}{371}}, \bibinfo{pages}{157} (\bibinfo{year}{1998}).

\bibitem[{\citenamefont{Faller and Schreck}(1997)}]{faller1997unsteady}
\bibinfo{author}{\bibfnamefont{W.~E.} \bibnamefont{Faller}} \bibnamefont{and}
  \bibinfo{author}{\bibfnamefont{S.~J.} \bibnamefont{Schreck}},
  {``}\bibinfo{title}{Unsteady fluid mechanics applications of neural
  networks},{''} \bibinfo{journal}{Journal of Aircraft}
  \textbf{\bibinfo{volume}{34}}, \bibinfo{pages}{48} (\bibinfo{year}{1997}).

\bibitem[{\citenamefont{Efe et~al.}(2004)\citenamefont{Efe, Debiasi, Ozbay, and
  Samimy}}]{efe2004modeling}
\bibinfo{author}{\bibfnamefont{M.~O.} \bibnamefont{Efe}},
  \bibinfo{author}{\bibfnamefont{M.}~\bibnamefont{Debiasi}},
  \bibinfo{author}{\bibfnamefont{H.}~\bibnamefont{Ozbay}}, \bibnamefont{and}
  \bibinfo{author}{\bibfnamefont{M.}~\bibnamefont{Samimy}}, in
  \emph{\bibinfo{booktitle}{Proceedings of the International Conference on
  Mechatronics, 2004.}} (\bibinfo{organization}{IEEE}, \bibinfo{year}{2004}),
  pp. \bibinfo{pages}{560--565}.

\bibitem[{\citenamefont{Lee et~al.}(1997)\citenamefont{Lee, Kim, Babcock, and
  Goodman}}]{lee1997application}
\bibinfo{author}{\bibfnamefont{C.}~\bibnamefont{Lee}},
  \bibinfo{author}{\bibfnamefont{J.}~\bibnamefont{Kim}},
  \bibinfo{author}{\bibfnamefont{D.}~\bibnamefont{Babcock}}, \bibnamefont{and}
  \bibinfo{author}{\bibfnamefont{R.}~\bibnamefont{Goodman}},
  {``}\bibinfo{title}{Application of neural networks to turbulence control for
  drag reduction},{''} \bibinfo{journal}{Phys. Fluids}
  \textbf{\bibinfo{volume}{9}}, \bibinfo{pages}{1740} (\bibinfo{year}{1997}).

\bibitem[{\citenamefont{Widrow et~al.}(1994)\citenamefont{Widrow, Rumelhart,
  and Lehr}}]{widrow1994neural}
\bibinfo{author}{\bibfnamefont{B.}~\bibnamefont{Widrow}},
  \bibinfo{author}{\bibfnamefont{D.~E.} \bibnamefont{Rumelhart}},
  \bibnamefont{and} \bibinfo{author}{\bibfnamefont{M.~A.} \bibnamefont{Lehr}},
  {``}\bibinfo{title}{Neural networks: applications in industry, business and
  science},{''} \bibinfo{journal}{Communications of the ACM}
  \textbf{\bibinfo{volume}{37}}, \bibinfo{pages}{93} (\bibinfo{year}{1994}).

\bibitem[{\citenamefont{Demuth et~al.}(2014)\citenamefont{Demuth, Beale,
  De~Jess, and Hagan}}]{demuth2014neural}
\bibinfo{author}{\bibfnamefont{H.~B.} \bibnamefont{Demuth}},
  \bibinfo{author}{\bibfnamefont{M.~H.} \bibnamefont{Beale}},
  \bibinfo{author}{\bibfnamefont{O.}~\bibnamefont{De~Jess}}, \bibnamefont{and}
  \bibinfo{author}{\bibfnamefont{M.~T.} \bibnamefont{Hagan}},
  \emph{\bibinfo{title}{Neural {N}etwork {D}esign}} (\bibinfo{publisher}{Martin
  Hagan}, \bibinfo{year}{2014}).

\bibitem[{\citenamefont{Raissi et~al.}(2017)\citenamefont{Raissi, Perdikaris,
  and Karniadakis}}]{raissi2017physics1}
\bibinfo{author}{\bibfnamefont{M.}~\bibnamefont{Raissi}},
  \bibinfo{author}{\bibfnamefont{P.}~\bibnamefont{Perdikaris}},
  \bibnamefont{and} \bibinfo{author}{\bibfnamefont{G.~E.}
  \bibnamefont{Karniadakis}}, {``}\bibinfo{title}{Physics Informed Deep
  Learning (Part I): Data-driven Solutions of Nonlinear Partial Differential
  Equations},{''} \bibinfo{journal}{arXiv preprint arXiv:1711.10561}
  (\bibinfo{year}{2017}).

\bibitem[{\citenamefont{Huang et~al.}(2006)\citenamefont{Huang, Zhu, and
  Siew}}]{huang2006extreme}
\bibinfo{author}{\bibfnamefont{G.-B.} \bibnamefont{Huang}},
  \bibinfo{author}{\bibfnamefont{Q.-Y.} \bibnamefont{Zhu}}, \bibnamefont{and}
  \bibinfo{author}{\bibfnamefont{C.-K.} \bibnamefont{Siew}},
  {``}\bibinfo{title}{Extreme learning machine: theory and applications},{''}
  \bibinfo{journal}{Neurocomputing} \textbf{\bibinfo{volume}{70}},
  \bibinfo{pages}{489} (\bibinfo{year}{2006}).

\bibitem[{\citenamefont{Cancelliere et~al.}(2017)\citenamefont{Cancelliere,
  Deluca, Gai, Gallinari, and Rubini}}]{cancelliere2017analysis}
\bibinfo{author}{\bibfnamefont{R.}~\bibnamefont{Cancelliere}},
  \bibinfo{author}{\bibfnamefont{R.}~\bibnamefont{Deluca}},
  \bibinfo{author}{\bibfnamefont{M.}~\bibnamefont{Gai}},
  \bibinfo{author}{\bibfnamefont{P.}~\bibnamefont{Gallinari}},
  \bibnamefont{and} \bibinfo{author}{\bibfnamefont{L.}~\bibnamefont{Rubini}},
  {``}\bibinfo{title}{An analysis of numerical issues in neural training by
  pseudoinversion},{''} \bibinfo{journal}{Computational and Applied
  Mathematics} \textbf{\bibinfo{volume}{36}}, \bibinfo{pages}{599}
  (\bibinfo{year}{2017}).

\bibitem[{\citenamefont{Majda and Wang}(2006)}]{majda2006nonlinear}
\bibinfo{author}{\bibfnamefont{A.}~\bibnamefont{Majda}} \bibnamefont{and}
  \bibinfo{author}{\bibfnamefont{X.}~\bibnamefont{Wang}},
  \emph{\bibinfo{title}{Nonlinear dynamics and statistical theories for basic
  geophysical flows}} (\bibinfo{publisher}{Cambridge University Press},
  \bibinfo{year}{2006}).

\bibitem[{\citenamefont{Selten}(1995)}]{selten1995efficient}
\bibinfo{author}{\bibfnamefont{F.~M.} \bibnamefont{Selten}},
  {``}\bibinfo{title}{An efficient description of the dynamics of barotropic
  flow},{''} \bibinfo{journal}{Journal of the Atmospheric Sciences}
  \textbf{\bibinfo{volume}{52}}, \bibinfo{pages}{915} (\bibinfo{year}{1995}).

\bibitem[{\citenamefont{Crommelin and Majda}(2004)}]{crommelin2004strategies}
\bibinfo{author}{\bibfnamefont{D.~T.} \bibnamefont{Crommelin}}
  \bibnamefont{and} \bibinfo{author}{\bibfnamefont{A.~J.} \bibnamefont{Majda}},
  {``}\bibinfo{title}{{Strategies for model reduction: comparing different
  optimal bases}},{''} \bibinfo{journal}{Journal of the Atmospheric Sciences}
  \textbf{\bibinfo{volume}{61}}, \bibinfo{pages}{2206} (\bibinfo{year}{2004}).

\bibitem[{\citenamefont{McWilliams}(2006)}]{mcwilliams2006fundamentals}
\bibinfo{author}{\bibfnamefont{J.~C.} \bibnamefont{McWilliams}},
  \emph{\bibinfo{title}{Fundamentals of geophysical fluid dynamics}}
  (\bibinfo{publisher}{Cambridge University Press}, \bibinfo{year}{2006}).

\bibitem[{\citenamefont{Ghil et~al.}(2008)\citenamefont{Ghil, Chekroun, and
  Simonnet}}]{ghil2008climate}
\bibinfo{author}{\bibfnamefont{M.}~\bibnamefont{Ghil}},
  \bibinfo{author}{\bibfnamefont{M.~D.} \bibnamefont{Chekroun}},
  \bibnamefont{and} \bibinfo{author}{\bibfnamefont{E.}~\bibnamefont{Simonnet}},
  {``}\bibinfo{title}{Climate dynamics and fluid mechanics: Natural variability
  and related uncertainties},{''} \bibinfo{journal}{Physica D: Nonlinear
  Phenomena} \textbf{\bibinfo{volume}{237}}, \bibinfo{pages}{2111}
  (\bibinfo{year}{2008}).

\bibitem[{\citenamefont{Lynch}(2008)}]{lynch2008origins}
\bibinfo{author}{\bibfnamefont{P.}~\bibnamefont{Lynch}},
  {``}\bibinfo{title}{The origins of computer weather prediction and climate
  modeling},{''} \bibinfo{journal}{Journal of Computational Physics}
  \textbf{\bibinfo{volume}{227}}, \bibinfo{pages}{3431} (\bibinfo{year}{2008}).

\bibitem[{\citenamefont{Greatbatch and Nadiga}(2000)}]{greatbatch2000four}
\bibinfo{author}{\bibfnamefont{R.~J.} \bibnamefont{Greatbatch}}
  \bibnamefont{and} \bibinfo{author}{\bibfnamefont{B.}~\bibnamefont{Nadiga}},
  {``}\bibinfo{title}{Four-gyre circulation in a barotropic model with
  double-gyre wind forcing},{''} \bibinfo{journal}{Journal of Physical
  Oceanography} \textbf{\bibinfo{volume}{30}}, \bibinfo{pages}{1461}
  (\bibinfo{year}{2000}).

\bibitem[{\citenamefont{San et~al.}(2011)\citenamefont{San, Staples, Wang, and
  Iliescu}}]{san2011approximate}
\bibinfo{author}{\bibfnamefont{O.}~\bibnamefont{San}},
  \bibinfo{author}{\bibfnamefont{A.~E.} \bibnamefont{Staples}},
  \bibinfo{author}{\bibfnamefont{Z.}~\bibnamefont{Wang}}, \bibnamefont{and}
  \bibinfo{author}{\bibfnamefont{T.}~\bibnamefont{Iliescu}},
  {``}\bibinfo{title}{Approximate deconvolution large eddy simulation of a
  barotropic ocean circulation model},{''} \bibinfo{journal}{Ocean Modelling}
  \textbf{\bibinfo{volume}{40}}, \bibinfo{pages}{120} (\bibinfo{year}{2011}).

\bibitem[{\citenamefont{Nadiga and Margolin}(2001)}]{nadiga2001dispersive}
\bibinfo{author}{\bibfnamefont{B.~T.} \bibnamefont{Nadiga}} \bibnamefont{and}
  \bibinfo{author}{\bibfnamefont{L.~G.} \bibnamefont{Margolin}},
  {``}\bibinfo{title}{Dispersive-dissipative eddy parameterization in a
  barotropic model},{''} \bibinfo{journal}{Journal of Physical Oceanography}
  \textbf{\bibinfo{volume}{31}}, \bibinfo{pages}{2525} (\bibinfo{year}{2001}).

\bibitem[{\citenamefont{Arakawa}(1966)}]{arakawa1966computational}
\bibinfo{author}{\bibfnamefont{A.}~\bibnamefont{Arakawa}},
  {``}\bibinfo{title}{Computational design for long-term numerical integration
  of the equations of fluid motion: \uppercase{T}wo-dimensional incompressible
  flow. \uppercase{P}art \uppercase{I}},{''} \bibinfo{journal}{Journal of
  Computational Physics} \textbf{\bibinfo{volume}{1}}, \bibinfo{pages}{119}
  (\bibinfo{year}{1966}).

\bibitem[{\citenamefont{Sirovich}(1987)}]{sirovich1987turbulence}
\bibinfo{author}{\bibfnamefont{L.}~\bibnamefont{Sirovich}},
  {``}\bibinfo{title}{{Turbulence and the dynamics of coherent structures. I.
  Coherent structures}},{''} \bibinfo{journal}{Quarterly of Applied
  Mathematics} \textbf{\bibinfo{volume}{45}}, \bibinfo{pages}{561}
  (\bibinfo{year}{1987}).

\bibitem[{\citenamefont{Ravindran}(2000)}]{ravindran2000reduced}
\bibinfo{author}{\bibfnamefont{S.}~\bibnamefont{Ravindran}},
  {``}\bibinfo{title}{A reduced-order approach for optimal control of fluids
  using proper orthogonal decomposition},{''} \bibinfo{journal}{International
  Journal for Numerical Methods in Fluids} \textbf{\bibinfo{volume}{34}},
  \bibinfo{pages}{425} (\bibinfo{year}{2000}).

\bibitem[{\citenamefont{Aubry et~al.}(1988)\citenamefont{Aubry, Holmes, Lumley,
  and Stone}}]{aubry1988dynamics}
\bibinfo{author}{\bibfnamefont{N.}~\bibnamefont{Aubry}},
  \bibinfo{author}{\bibfnamefont{P.}~\bibnamefont{Holmes}},
  \bibinfo{author}{\bibfnamefont{J.~L.} \bibnamefont{Lumley}},
  \bibnamefont{and} \bibinfo{author}{\bibfnamefont{E.}~\bibnamefont{Stone}},
  {``}\bibinfo{title}{The dynamics of coherent structures in the wall region of
  a turbulent boundary layer},{''} \bibinfo{journal}{Journal of Fluid
  Mechanics} \textbf{\bibinfo{volume}{192}}, \bibinfo{pages}{115}
  (\bibinfo{year}{1988}).

\bibitem[{\citenamefont{Goodfellow et~al.}(2016)\citenamefont{Goodfellow,
  Bengio, Courville, and Bengio}}]{goodfellow2016deep}
\bibinfo{author}{\bibfnamefont{I.}~\bibnamefont{Goodfellow}},
  \bibinfo{author}{\bibfnamefont{Y.}~\bibnamefont{Bengio}},
  \bibinfo{author}{\bibfnamefont{A.}~\bibnamefont{Courville}},
  \bibnamefont{and} \bibinfo{author}{\bibfnamefont{Y.}~\bibnamefont{Bengio}},
  \emph{\bibinfo{title}{Deep learning}}, vol.~\bibinfo{volume}{1}
  (\bibinfo{publisher}{MIT press Cambridge}, \bibinfo{year}{2016}).

\bibitem[{\citenamefont{Carrillo et~al.}(2016)\citenamefont{Carrillo, Que, and
  Gonz{\'a}lez}}]{carrillo2016estimation}
\bibinfo{author}{\bibfnamefont{M.}~\bibnamefont{Carrillo}},
  \bibinfo{author}{\bibfnamefont{U.}~\bibnamefont{Que}}, \bibnamefont{and}
  \bibinfo{author}{\bibfnamefont{J.~A.} \bibnamefont{Gonz{\'a}lez}},
  {``}\bibinfo{title}{{Estimation of Reynolds number for flows around cylinders
  with lattice Boltzmann methods and artificial neural networks}},{''}
  \bibinfo{journal}{Physical Review E} \textbf{\bibinfo{volume}{94}},
  \bibinfo{pages}{063304} (\bibinfo{year}{2016}).

\end{thebibliography}
\bibliographystyle{apsrevlong}

\end{document}